# Efficient prediction of highly anisotropic excitonic properties in the layered antiferromagnet CrSBr via time-dependent density functional theory

**Ashwin Ramasubramaniam,**[1,2,*] **Daniel Hernangómez-Pérez,**[3] **Javier Junquera,**[4] **María Camarasa-Gómez**[5]

[1] Department of Mechanical and Industrial Engineering, University of Massachusetts, Amherst, MA 01003, U.S.A.

[2] Graduate Program in Materials Science and Engineering, University of Massachusetts, Amherst, MA 01003, U.S.A.

[3] CIC nanoGUNE BRTA, Tolosa Hiribidea 76, 20018 San Sebastián, Spain

[4] Departamento de Ciencias de la Tierra y Física de la Materia Condensada, Facultad de Ciencias, Universidad de Cantabria, Avenida de los Castros s/n, E-39005 Santander, Spain

[5] Centro de Física de Materiales (CFM-MPC), CSIC-UPV/EHU, Paseo Manuel de Lardizabal 5 20018 Donostia-San Sebastián, Spain

## ABSTRACT

CrSBr, a layered anisotropic van der Waals antiferromagnet, has recently emerged as a versatile platform where strong coupling between optical excitations and magnetic order enables magneto-optical control in low dimensions. While experiments have progressed rapidly, predictive and reliable *ab initio* descriptions remain limited to self-consistent, many-body perturbation theory that is computationally expensive and technically challenging. Here we present an alternative approach that accurately predicts the electronic and optical properties of CrSBr at substantially lower computational cost, while retaining quantitative accuracy in the coupling between excitons and magnetic order. Using a tuned hybrid density functional with on-site corrections, we reproduce fundamental and optical gaps and quantitatively capture the interaction between excitonic transitions and magnetic order. We then employ this functional to investigate excitonic shifts induced by spin canting, that would result from applying an external magnetic field. Our results establish an efficient framework for modeling excitonic and magneto-optical phenomena in layered magnetic semiconductors.

**Corresponding Author**

* ashwin@engin.umass.edu

**TOC Graphic**

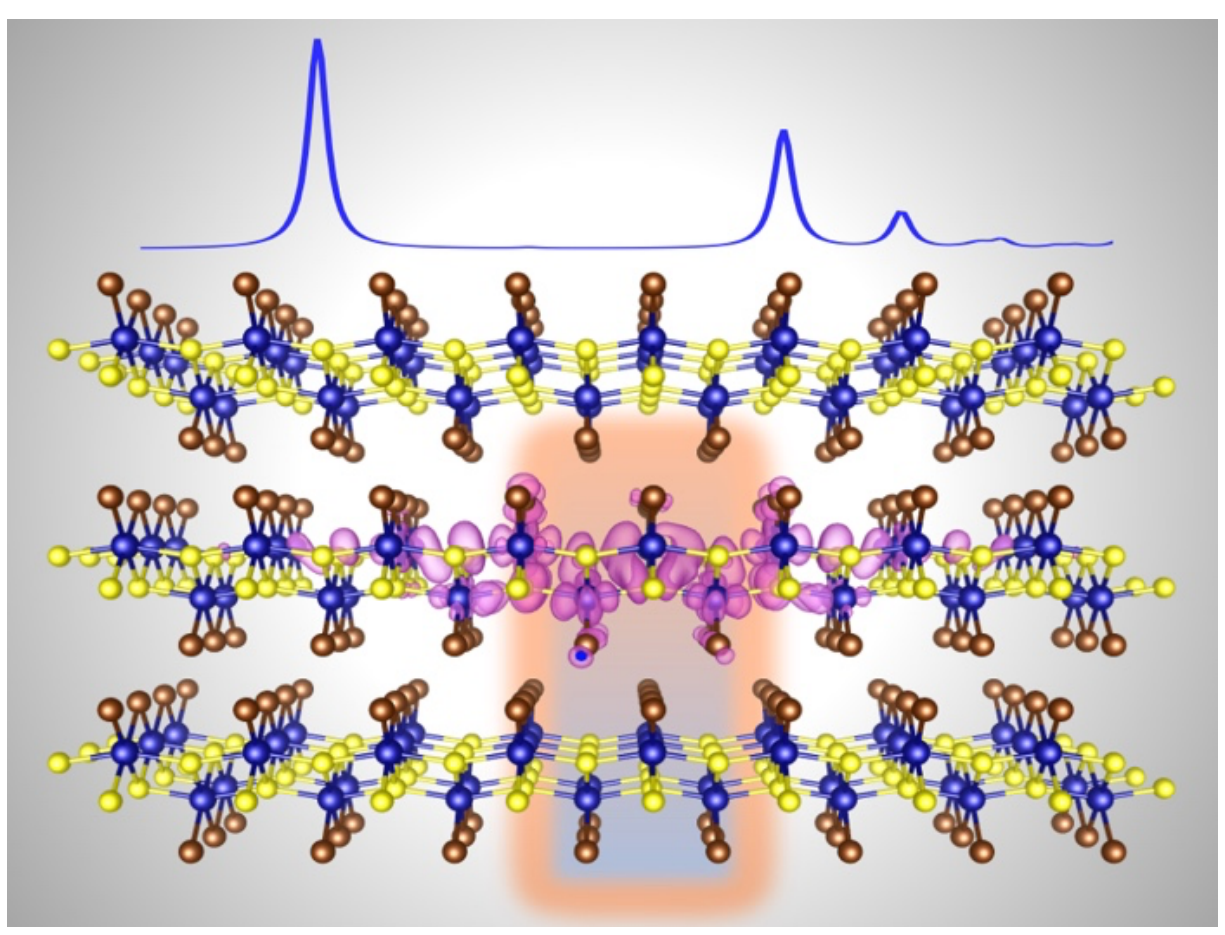

Layered van der Waals materials have been attracting significant research interest over the past two decades, driven by discoveries of new materials,[1,2] physical phenomena,[3,4] and technological advancements.[5–7] More recently, layered van der Waals magnets have emerged as a particularly active area of research, where the unique nature of low-dimensional quasiparticle excitations, arising from quantum confinement, couple strongly to magnetic order, leading to a variety of new emergent phenomena.[8–15] Among layered van der Waals magnets, CrSBr – an A-type antiferromagnetic semiconductor – has attracted intense scrutiny due to several intriguing properties, including strong coupling of excitons and magnons,[16–18] strain- and gate-tunability of excitons,[15,19] magnetically dressed exciton-polaritons,[20–22] and defect-induced magnetic phase transitions,[23] to name a few. Additionally, CrSBr is relatively stable at ambient conditions and possesses a relatively high Néel temperature,[24] making it attractive for applications in magnonics and spintronics.

In its bulk form, CrSBr possesses an orthorhombic structure (space group Pmmn; no. 59).[25] Below the Néel temperature (~132 K [24]) the stacked 2D layers couple antiferromagnetically and, within each layer, the spins align along the *b*-axis (Fig. 1a), which constitutes the easy-axis of this material. ARPES and vertex-corrected quasiparticle self-consistent GW ($QSG\widehat{W}$) calculations reveal that the band edges are dominated by Cr *d*-states with some degree of mixing from the S and Br *p*-states.[26,27] The low energy optical spectrum is dominated by Cr *d*-*d* transitions with two well characterized bright excitons located at ~1.37 eV ($X_A$)[28] and ~1.76 eV ($X_B$).[29] Both excitons are highly anisotropic, being extended along the *b*-axis but largely confined along the material *a*-axis, which is attributed to the quasi-1D nature of charge localization in this material.[28,30,31] By contrast, the fundamental band gap of bulk CrSBr is less well established, although ARPES studies place a lower bound of ~1.85 eV on this value while self-consistent GW calculations place this value closer to 1.95 eV [31] or 2.05 eV.[26] These results indicate that, at least, the $X_A$ exciton is strongly bound even in the bulk material, which is a departure from the norm for other layered van der Waals semiconductors where such large exciton binding typically only becomes apparent in the 2D limit.[32–35] This unusual behavior has been attributed to the anisotropic quasi-1D nature of the charge distribution that results in reduced dielectric screening, with additional localization resulting from the antiferromagnetic ordering between layers that suppresses interlayer coupling.[16]

While experimental research on CrSBr has proceeded at a rapid pace along several fronts, accurate first-principles modeling remains comparatively more limited and based primarily on many-body perturbation theory (MBPT). Specifically, computationally expensive self-consistent GW calculations, and in particular quasi-particle self-consistent GW ($QSG\widehat{W}$) calculations, which do not suffer from starting-point dependencies.[26,27,31,32,36] have yielded electronic bandstructures and optical absorption spectra for bulk CrSBr that are in excellent quantitative agreement with several experimental studies.[26,27,32,36] However, the computational cost of these self-consistent methods limits their applicability to larger or more complex systems, prompting the development of alternative first-principles approaches such as semilocal density functionals,[37,38] hybrid density functionals,[28,39] DFT+$U$,[40,41] or single-shot GW[28,42,43]. While these methods capture the semiconducting nature of CrSBr, the calculated fundamental gaps are strongly and consistently underestimated relative to ARPES measurements. Moreover, the optical absorption spectra calculated via the Bethe-Salpeter equation (BSE) subsequent to single-shot GW,[28,42,43] only capture the $X_A$ exciton energy correctly, largely missing the optical features at higher energies (such as the $X_B$ exciton and satellite peaks). Other studies have adopted approaches based on dynamical mean-field theory [44] or cluster models [45,46] to address the correlated nature of the Cr 3*d* orbitals. However, they focus on understanding specific aspects of Cr *d*-*d* transitions and Mott localization and therefore, do not account more broadly for the complex magneto-optical and excitonic properties that dictate the performance and functionality of CrSBr. Collectively, these theoretical efforts raise a fundamental question: can the electronic structure and optical response of bulk CrSBr be described with *quantitative* accuracy within the framework of generalized Kohn-Sham (GKS) time-dependent density functional theory (TDDFT) or are MBPT and/or quantum chemistry-based methods unavoidable? Indeed, if the conventional machinery of TDDFT could be reliably applied for CrSBr (and, possibly, similar magnetic semiconductors), this would significantly reduce the computational cost and enable routine calculations within widely used first-principles frameworks.

The purpose of this paper is thus to demonstrate that a simple tuned hybrid density functional can capture the key electronic and optical features of bulk CrSBr, as well as their coupling to magnetic order, with *high quantitative accuracy*. Specifically, we show that by tuning a hybrid functional to capture the $X_A$ and $X_B$ exciton energies of bulk CrSBr, we obtain excellent qualitative agreement for the nature of the band edges as well as quantitative estimates for the fundamental gap, in agreement with ARPES and $QSG\widehat{W}$ many-body results. Using this tuned functional, we further investigate the coupling between excitons and the magnetic order, and obtaining quantitatively accurate estimates for the excitonic shifts of the $X_A$ and $X_B$ peaks, as well as changes in oscillator strengths and emergence of satellite features in the many-body optical spectrum due to canting of the magnetic moments. More broadly, our results demonstrate that GKS-TDDFT provides an efficient and predictive framework for modeling excitonic and magneto-optical phenomena in CrSBr and related layered magnetic semiconductors, complementing and simplifying existing state-of-the-art first-principles approaches.

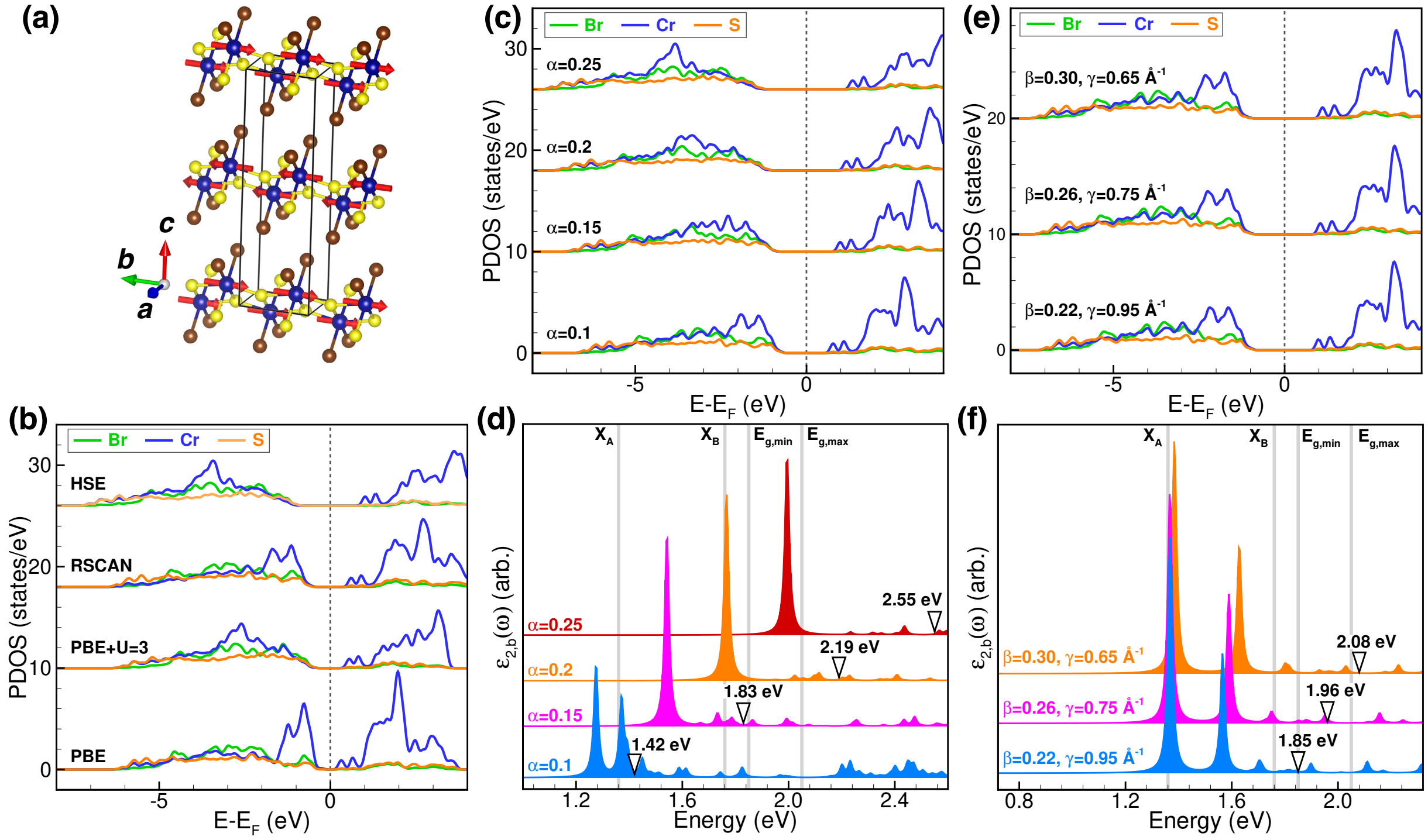

**Figure 1:** (**a**) Crystal structure of bulk CrSBr; Cr, S, Br are indicated by blue, yellow, and brown spheres, respectively, and the magnetic moments (AFM state) are indicated by red arrows. (**b**) Species-wise projected density of states (PDOS) using selected density functionals. (**c**) Species-wise PDOS using global hybrids with varying fractions of exact exchange and (**d**) their calculated optical spectra (imaginary part of the dielectric function) along the *b*-axis. (**e, f**) Same as (c, d) but with range-separated hybrids consisting of 100% semilocal exchange (PBE) in the short range and a fraction $\beta$ of exact-exchange in the long range, $\gamma$ being the range-separation parameter. In (d) and (f), we have marked by grey vertical lines the reference energies of the $X_A$ (1.36 eV) and $X_B$ (1.76 eV) excitons as well as the approximate bounds on the fundamental gap ($E_{g,min}$=1.85 eV and $E_{g,max}$=2.05 eV); the fundamental gaps are indicated by inverted triangles. The Fermi level is set mid-gap in all PDOS plots and only the majority spin PDOS are indicated, the minority spin states being degenerate.

We begin by examining how different density functionals affect the electronic structure of CrSBr, in particular, focusing on how the character of the band edges, fundamental gaps, and optical absorption spectra evolve with varying treatments of the exchange interaction. Figure 1(b) displays the species-wise projected density of states for two semilocal functionals – PBE [47] and the RSCAN metaGGA[48] – as well as PBE+$U$ and the HSE [49] functional (that only contains short-range exact exchange). Both PBE and RSCAN predict that the valence band edge is dominated by Cr (*d*- states) with some mixing from S and Br states, in agreement with $QSG\widehat{W}$ and DMFT calculations.[26,44] In contrast, both PBE+$U$ (for which we use $U$=3 eV, as an example) and HSE shift the Cr states to lower energies, leading to much greater mixing of anion $p$- and cation $d$-states at the valence band edge. The similar behavior of PBE+$U$ and HSE can be understood by recognizing that the $U$ correction effectively provides a simplified screened Fock-like exchange in the short-range,[50–52] the difference being that the $U$ correction is applied only to a selected subspace (here, Cr $d$-orbitals) whereas in HSE the short-range exact exchange is applied to all orbitals. In all cases though, the conduction band edge is predominantly of Cr character with the clear presence of two peaks at the band edge that correspond to the well-known flat conduction bands along the $\Gamma - X$ direction (crystallographic $a$ axis).[28] These results provide the first useful hint that the inclusion of short-range exact exchange can, perhaps unexpectedly, lead to a qualitatively incorrect description of the valence band edge. However, we cannot yet determine the impact on the excitonic properties, as these functionals lack the proper long range behavior required for obtaining bound excitons.[53,54]

Thus, we consider next hybrid functionals that contain long-range exact exchange, beginning with global hybrids where the exchange potential, $v_x$, is given by

$$v_x = \alpha v_{XX} + (1 - \alpha) v_{SL}, \tag{1}$$

where $\alpha$ is the fraction of exact exchange (XX) and $1 - \alpha$ is the fraction of semilocal (SL) exchange. We use PBE semilocal exchange and correlation throughout this work. Figure 1(c) displays the PDOS for fractions of exact exchange ranging from 10–25% from which we once again observe that with increasing fraction of exact exchange the Cr states shift progressively deeper into the valence band, which is undesirable. The fundamental gap increases linearly with increasing exact exchange (Fig. S1). The optical absorption spectra are displayed in Figure 1(d) from which we note two significant features: first, the absorption onset (defined as the energy of the first excitonic peak) is blue-shifted with increasing exact exchange and, second, the satellite peaks that are 0.2–0.3 eV above the first peak are correspondingly rapidly suppressed. The blue shift of the first exciton peak is again nearly linear correlated with the fraction of exact exchange (Fig. S1) although at a slower rate than the fundamental gap, leading to stronger exciton binding with increasing exact exchange. Looking specifically at the case of $\alpha = 0.15$, for which the fundamental gap (1.83 eV) is nearly at the lower bound (1.85 eV), it is already apparent that the first excitonic peak is substantially higher (~0.2 eV) than that of the $X_A$ exciton and, moreover, the spectral features in the vicinity of the $X_B$ exciton range are greatly suppressed.

Clearly, while global hybrids are unable to describe bulk CrSBr correctly, the earlier observation that semilocal functionals appear to capture the qualitative character of the valence band edge correctly suggest the possibility of employing a screened range-separated hybrid (SRSH) functional [34,35,55–57] wherein short-range exact exchange is minimized while a fraction of long-range exact exchange is retained to ensure correct dielectric screening. As an extreme case, we consider one such form for an SRSH where the exchange potential, $v_x$, is given by

$$v_x = v_{SL}^{SR} + \beta v_{XX}^{LR} + (1 - \beta) v_{SL}^{LR}, \tag{2}$$

such that the short-range (SR) component is purely semilocal, switching to a long-range (LR) component that admixes a fraction, $\beta$, of exact exchange with a complementary fraction, $1 - \beta$, of semilocal (PBE) exchange. Treating $\beta$ and the range-separation parameter, $\gamma$, as tunable parameters, we calculated the electronic structure and optical spectra with the aim of finding suitable $\beta - \gamma$ pairs for which the $X_A$ exciton energy is close to the reference value while the fundamental gap is within the acceptable bounds. Figures 1(e, f) show the outcomes of these calculations for a few select cases from which it is clear that suppressing short-range exact exchange indeed produces band edges that are predominantly composed of Cr states with low mixing of S and Br states. It is also clear that the energy of the $X_B$ exciton is much lower than desired even though the oscillator strength is significantly higher than with pure global hybrids. These conclusions also hold good for more general cases where a small fraction of exact exchange was included in the short-range (see Fig. S2 for an example). Altogether, these failures do provide one systematic and key insight: while exact-exchange is generally beneficial for mitigating self-interaction errors, it is actually detrimental in the present situation for the Cr $d$-orbitals, causing *over-localization*. Put differently, we need to consider *orbital-wise* corrections and treat the $s$- and $p$-states of the anions differently from the correlated Cr $d$-orbitals, motivating an orbital-selective approach in which we combine a (global) hybrid functional with on-site corrections – the so-called "hybrid+$V_w$" method.[58–60]

The hybrid+$V_w$ method was proposed by Ivády *et al.*[58–60] for alleviating inaccuracies that arise in correlated systems from treating localized and extended states on the same footing vis-à-vis screening. Specifically, they showed that within the fully-localized limit, hybrid DFT can be viewed as analogous to DFT+$U$ where the strength of the on-site potential for the former is now directly proportional to the fraction of exact exchange, $\alpha$. Viewed in this light, it is not surprising that the homogeneous screening provided by a single value of $\alpha$ might not be appropriate for localized versus extended states, much as different states require different $U$ corrections in the DFT+$U$ method. Thus, Ivády *et al.* augmented the global hybrid potential (Eq. 1) with an additional on-site potential of the form [59]

$$V_m^\sigma = V_w\left(\frac{1}{2} - n_m^\sigma\right), \quad (3)$$

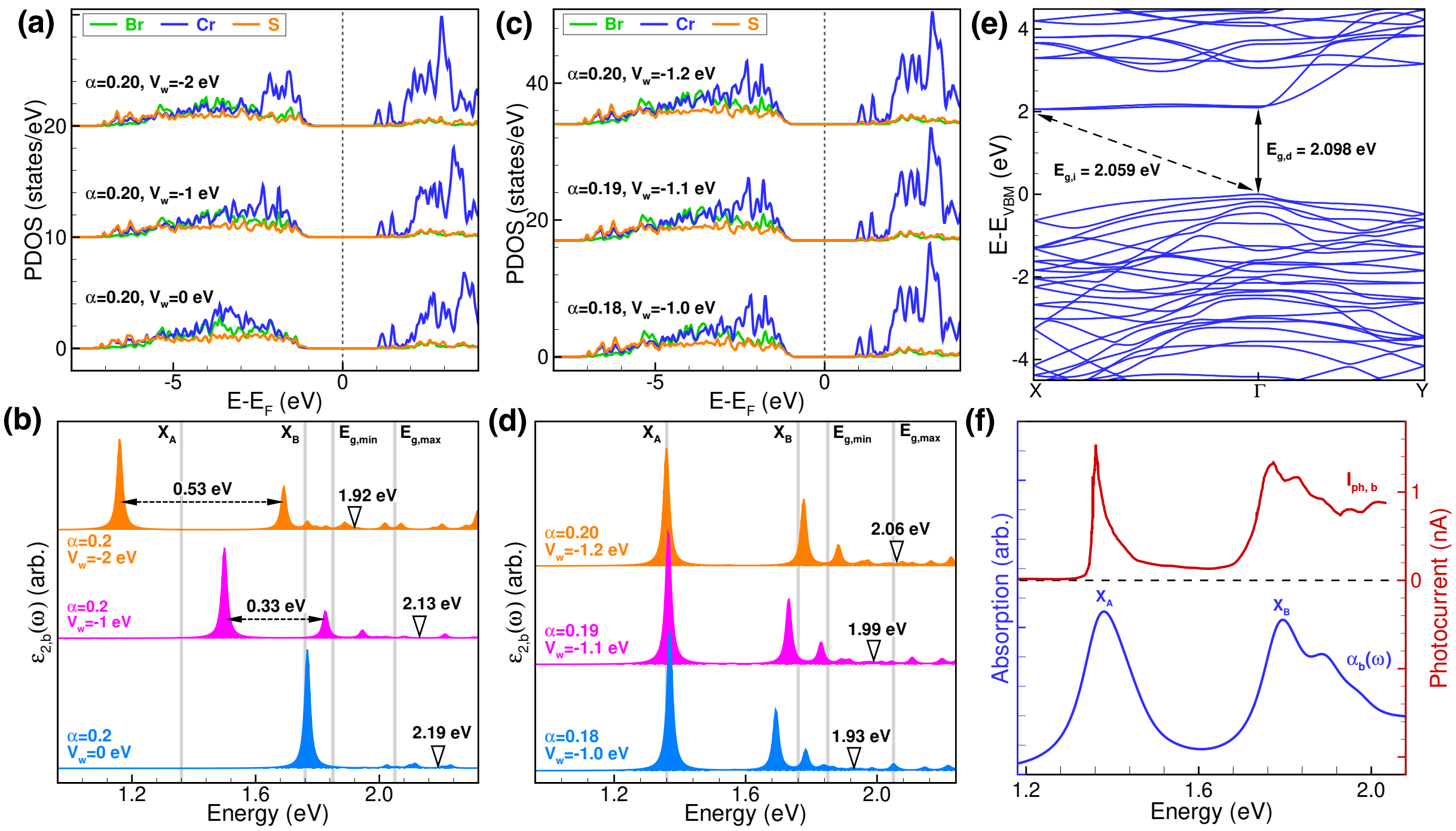

**Figure 2:** (**a**) Species-wise PDOS of bulk CrSBr using the hybrid+$V_w$ approach and (**b**) the corresponding optical spectra (imaginary part of the dielectric function) along the $b$-axis. (**c, d**) Same as (a, b) but with the inclusion of spin-orbit coupling, and (**e**) the Wannier-interpolated bandstructure for the case of $\alpha = 0.2$ and $V_w = -1.2$ eV. The fundamental gap is weakly indirect between the $\Gamma$ and X points. (**f**) Comparison of the photocurrent spectrum ($I_{ph,b}$) measured for incident light polarized along the $b$-axis, extracted from Ref. 62, and the calculated absorption spectrum [$\alpha_b(\omega) \propto \omega\kappa_b(\omega)$; $\kappa_b(\omega)$ is the extinction coefficient] using $\alpha = 0.2$ and $V_w = -1.2$ eV, and a broadening of 50 meV. In (b) and (d), we have marked by grey vertical lines the reference energies of the $X_A$ (1.36 eV) and $X_B$ (1.76 eV) excitons as well as the approximate bounds on the fundamental gap ($E_{g,min}$=1.85 eV and $E_{g,max}$=2.05 eV); the fundamental gaps are indicated by inverted triangles. The Fermi level is set mid-gap in all PDOS plots and only the majority spin PDOS are indicated in (a), the minority spin states being degenerate.

that is applied selectively to a set of localized atomic-like orbitals $\varphi_m^\sigma$, $V_w$ being the strength of the on-site potential and $n_m^\sigma$ being the on-site occupation matrix. The form of this potential is motivated by the simplified rotationally-invariant form of DFT+$U$ proposed by Dudarev *et al.*[61] Note that the parameter $V_w$ can take on either positive or negative values and can be determined, in principle, by satisfying the ionization potential theorem, e.g., for defect states of transition-metal dopants in semiconductors.[58,59] Within this framework, $V_w$ can also be interpreted as a static, orbital-dependent self-energy correction that accounts for the discrepancy between the hybrid exchange-correlation potential and the true many-body screened interaction. By targeting a specific orbital subspace, it provides a localized adjustment to the microscopic screening while leaving the fundamental band gap essentially unperturbed. For the present, we treat $\alpha$ and $V_w$ as adjustable parameters that are tuned to obtain the desired energies of the $X_A$ and $X_B$ excitons while fulfilling the bounds on the fundamental gap.

Figure 2(a) displays the outcome of the hybrid+$V_w$ method, applied to bulk CrSBr (AFM phase), where we fix the fraction of exact exchange ($\alpha = 0.20$) and vary the onsite potential, $V_w$. As evident, applying negative values of $V_w$ to the Cr $d$-orbitals systematically cancels out the excess exact exchange acting on these orbitals, shifts these states progressively towards the valence band edge, reduces the extent of *p*-$d$ mixing (qualitatively in line with $QSG\widehat{W}$ and DMFT calculations), and reduces the fundamental band gap. Moreover, as seen from Figure 2(b), the $X_A$ excitonic peak shift to lower energies while the $X_B$ excitonic peak acquires more intensity as the Cr character of the valence edges increases. A more complete discussion of the dependence of the fundamental gap and the $X_A$ and $X_B$ exciton energies on $\alpha$ and $V_w$ is provided in Section S3 of the Supporting Information. Most interestingly, it is now possible to control the splitting between the $X_A$ and $X_B$ excitons, unlike with SRSH functionals where the splitting remains in the range of 0.2–0.3 eV if the fundamental gap is to stay within acceptable bounds. Based on this insight, we systematically tuned $\alpha$ and $V_w$ – now with full inclusion of spin-orbit coupling and Cr magnetic moments correctly aligned (antiferromagnetically) along the easy $b$-axis – and the outcome of these calculations is displayed for select cases in Figure 2(c) and (d). Specifically, in the tuning procedure, we demand first that the energy of the $X_A$ excitonic peak is as close as possible to the correct reference energy and then further require that the $X_B$ excitonic peak is also close to its reference value while ensuring that the fundamental gap stays within acceptable bounds. The choice of parameters is, of course, not unique but we choose $\alpha = 0.2$ and $V_W = -1.2$ eV as a representative case for now. Figure 2(e) displays the electronic bandstructure from which we note the nearly flat pair of bands along the $\Gamma - X$ direction [16,28] at the conduction band edge and we also find the fundamental gap to be weakly indirect. Figure 2(f) displays a comparison between the photocurrent spectrum for light polarized along the $b$-axis, reported in Ref. 62, and our calculated absorption coefficient for the same direction, which we find to be in excellent agreement, both with respect to peak positions and intensities.

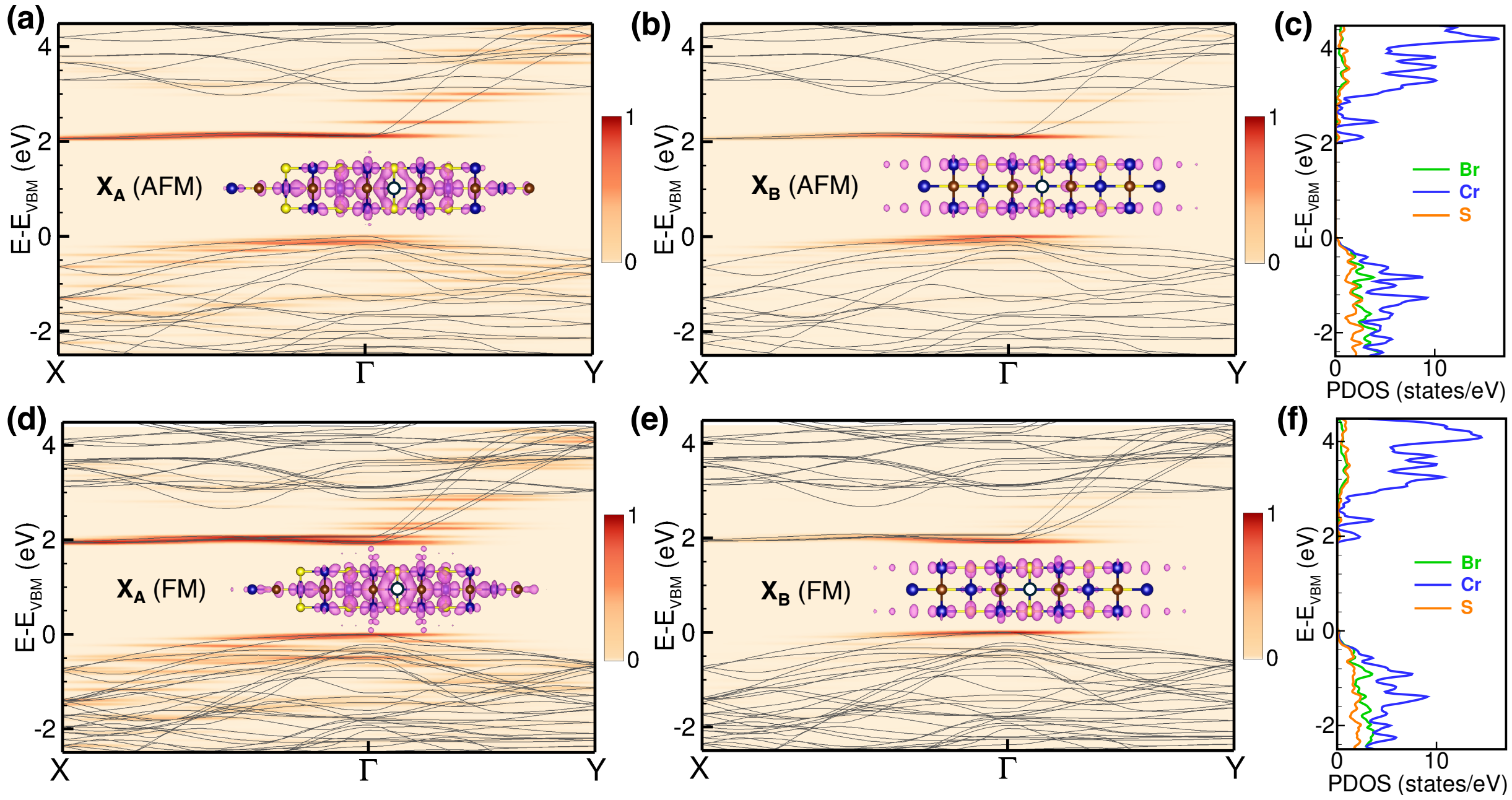

**Figure 3:** (**a**, **b**) Wannier-interpolated bandstructure of bulk CrSBr (AFM phase) with superposed heatmaps indicating the contributions from the bands to the $X_A$ and $X_B$ excitons. Insets depict the real-space charge density of the two excitons, with the hole position being fixed in the vicinity of a Cr atom (white circle). (**c**) Species-wise PDOS for the bulk AFM phase. (**c, d, e**) Same as (a, b, c) but for the bulk ferromagnetic (c-FM) phase with Cr magnetic moments aligned along the $c$-axis. All calculations are performed with $\alpha = 0.2$ and $V_w = -1.2$ eV, and with the inclusion of spin-orbit coupling.

To gain further insight into the composition and nature of the excitons, we display in Figures 3(a, b) the bandstructures of the bulk AFM phase and indicate via heatmaps (derived from BSE fatbands, as explained in Sec. S4) the contributions of the bands to the $X_A$ and $X_B$ excitons, respectively. Unlike the $X_B$ exciton that is composed largely of band-edge states, the $X_A$ exciton has more contributions from states deeper in the valence and conduction bands and is also more spread out in reciprocal space. The tighter-bound $X_A$ exciton is more localized in real space than the $X_B$ exciton and both are strongly anisotropic, being elongated along the *b*-axis and confined along the *a*-axis. Both excitons have large Cr *d-d* contributions with some *p-d* character, as seen from their charge density distributions. We are unable at present to disentangle further the onsite vs. intersite nature of the *d-d* excitations.[36] With the optimal parameters, we also estimate the exciton binding energy (difference between the fundamental gap and the exciton energy) to be ~0.7 eV for the $X_A$ exciton and ~0.3 eV for the $X_B$ exciton. These results are in very good agreement with the state-of-the-art calculations reported in Ref. 36 and recently reported trARPES measurements (~0.8 eV for the $X_A$ exciton).[63]

Up to this point, we have demonstrated that the hybrid+$V_w$ functional can be tuned to reproduce a couple of known targets and, subsequently, produce bandstructures and absorption spectra that agree well with the literature. A stringent test of the approach is its ability to capture magnetically driven excitonic shifts that are well characterized in experiments.[29,36,42] Figure 4(a) displays the calculated optical spectra as a function of canting of Cr spins out of plane, effectively mimicking a gradually increasing external magnetic field (applied along the $c$-axis) that causes spins to eventually align ferromagnetically along the $c$-axis (see Computational Methods). Two features are particular noteworthy from these calculations: first, the $X_A$ exciton redshifts by 38 meV whereas the $X_B$ exciton redshifts more substantially by 132 meV; second, there is a progressive redistribution of spectral weight between the $X_B$ exciton and its satellite peaks (originally in the ~1.8–2.0 eV range) leading ultimately to three peaks ($X_B, X_B', X_B''$ [36]) of similar intensities in the FM phase. The  fundamental gap of the c-FM phase is calculated to be 1.92 eV, a decrease of 0.14 eV relative to the AFM phase, which is in excellent agreement $QSG\widehat{W}$ results.[36] This reduction in band gap occurs as interlayer hopping is no longer spin-forbidden for the c-FM case, unlike the AFM case.[16,29] The redshift of the $X_B$ exciton then nearly tracks the change in fundamental gap, being composed mostly of band edge states (Fig. 3e), whereas the more-localized $X_A$ exciton, which is composed of deeper states (Fig. 3d), is less affected by overall changes in the fundamental gap. This is also consistent with Śmiertka *et al.*'s [36] findings, which they attribute to the $X_A$ exciton having larger contributions from onsite *d-d* transitions, making them less sensitive to band-structure changes. Overall, the calculated redshifts of the $X_A$ and $X_B$ excitons, while slightly higher than measured ones ($X_A$: 20 meV,[42] 14±2 meV,[21] ~12 meV [36]; $X_B$: 110 meV,[36] ~100 meV [29]), are nevertheless in reasonable agreement. Finally, in Figure 4(b) we also display the real and imaginary parts of $\varepsilon_{ab}$, which is the only non-zero off-diagonal component of the dielectric tensor in the spin-canted phases where time-reversal symmetry is broken. Strong resonances in $\varepsilon_{ab}$ are visible at the $X_A$ and $X_B$ exciton frequencies and the strength of these resonances is clearly proportional to the exciton oscillator strengths. We expect that these resonant features will also be manifested in magneto-optical (e.g., MOKE, MCD) or magneto-transport (e.g., AHE) measurements.

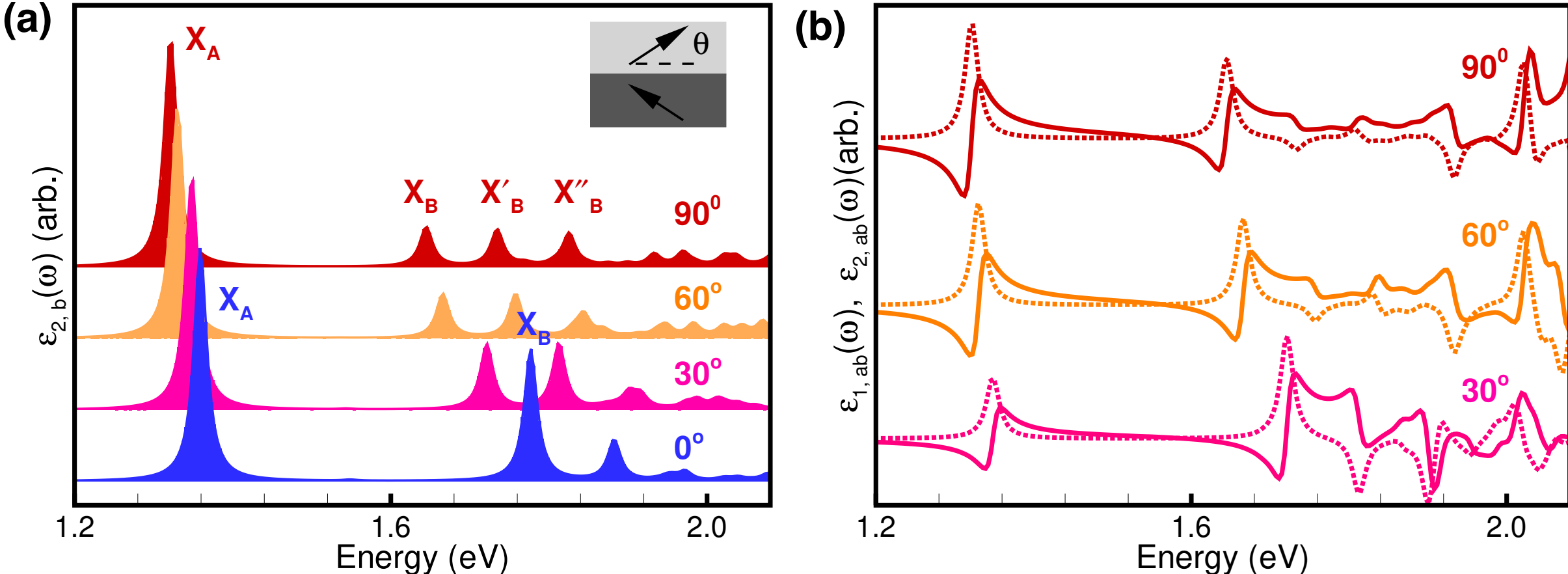

**Figure 4: (a)** Optical spectra (imaginary part of the dielectric function) along the *b*-axis as a function of orientation of Cr spins. Starting from the blue spectrum (AFM case) the Cr spins progressively cant out of plane (in steps of 30°) until they are completely ferromagnetically aligned normal to the layers (along the *c*-axis). **(b)** Real (dashed) and imaginary (solid) parts of the single non-zero off-diagonal component $\varepsilon_{ab}$ of the dielectric tensor for various canting angles; $\varepsilon_{ab}$ is trivially zero for the AFM case. All calculations are performed with $\alpha = 0.2$ and $V_w = -1.2$ eV, and with the inclusion of spin-orbit coupling.

To conclude, we have presented a simple yet effective *ab initio* approach for calculating the electronic and excitonic properties of bulk CrSBr with quantitative accuracy. By combining a global hybrid functional with orbital-specific on-site corrections, the approach reproduces key excitonic and magnetic responses at substantially lower computational cost than MBPT, while remaining fully variational and compatible with standard first-principles codes. In particular, with only two tunable parameters, the functional reproduces fundamental and optical gaps, providing a predictive framework for CrSBr and potentially related layered magnetic semiconductors. Therefore, this approach will enable efficient calculations of spin-dependent properties such as magnetocrystalline anisotropy and exchange couplings, including many-body effects at the TDDFT level, facilitating future studies of spin dynamics. Although empirical tuning is required, parameters could in the future be determined using optimal-tuning procedures that have been developed recently for the solid state.[35,55,64–66] Lastly, a natural next step is to explore the applicability of the hybrid+$V_w$ approach across weakly and strongly correlated systems, and its utility as a starting point for DMFT.

## COMPUTATIONAL METHODS

All calculations were performed using the Vienna *Ab Initio* Simulation Package (VASP).[67,68] Standard projector-augmented-wave (PAW) potentials [69] provided with VASP were used to model core and valence electrons. The valence configurations of Cr, S, Br are $3p^6 4s^1 3d^5$, $3s^2 3p^4$, and $4s^2 4p^5$, respectively. Semilocal contributions to electron exchange and correlation were modeled using the Perdew-Burke-Ernzerhof (PBE)[47] form of the generalized-gradient approximation in correspondence with the PAW potentials. The kinetic energy cutoff was set to 550 eV, and the Brillouin zone was sampled using a $10 \times 8 \times 2$ $\Gamma$-centered $k$-point mesh. A Gaussian smearing of 50 meV was used in all calculations. Linear-response TDDFT calculations [70] were performed in the Tamm-Dancoff approximation,[71] using 60 occupied and 20 unoccupied bands (as determined from convergence tests). Gaussian broadening of 10 meV was used for all optical spectra except when comparing to experiments (Fig. 2f) where 50 meV broadening was used to facilitate better comparison. Calculations with canted spins were performed using the constrained magnetization implementation in VASP and only the orientations of the spins were fixed at desired canting angles. SRSH calculations were performed using our own custom implementation, while hybrid+$V_w$ calculations were performed by merging hybrid DFT and DFT+$U$ (rotationally-invariant form of Dudarev *et al.*[61]) tags in the VASP input file. Atomic positions and lattice parameters were fixed at their experimental values [25] (see Sec. S2). Selected tests, verifying the hybrid+$V_w$ approach with the SIESTA package, are provided in Section S5.

**ACKNOWLEDGEMENTS**

A.R. thanks Dr. Swagata Acharya and Prof. Leeor Kronik for several insightful discussions. A.R. also gratefully acknowledges computational support from the Office of Information Technology at the University of Massachusetts Amherst and the Massachusetts Green High-Performance Computing Center. D.H.-P. is grateful for funding from the Diputación Foral de Gipuzkoa through Grants 2023-FELL-000002-01, 2024-FELL-000009-01, and 2025-FELL-000004-01 and Red Guipuzcoana R&D Project TRAFIC (Project 2025-CIE4-000036-01). D.H.-P. is also grateful for the support of the Spanish MICIU/AEI/10.13039/501100011033 and ERDF/EU, through grant PID2023-147324NA-I00 and from IKUR Strategy, Quantum Technologies 2025 project M-Twist, from the Department of Science, Universities and Innovation of the Basque Government. J.J. acknowledges financial support from grant no. PID2022-139776NBC63 funded by MICIU/AEI/10.13039/501100011033 and ERDF/EU/AEI/10.13039/ 501100011033 and by ERDF/EU "A way of making Europe" by the European Union. M.C.-G. acknowledges support from grant no. PID2024-159869NA-I00 funded by MICIU/AEI/10.13039/501100011033 and ERDF/EU, and from the Diputación Foral de Gipuzkoa through grants 2024-FELL-000007-01 and 2025-FELL-000009-01. D.H.-P. and M.C.-G. thankfully acknowledge RES resources provided by the Barcelona Supercomputing Center in MareNostrum 5 to FI-2024-2-0023, FI-2024-3-0024.

## SUPPORTING INFORMATION

Supporting figures for global and range-separated hybrids; unit cell and atomic coordinates of bulk CrSBr; parameter tuning plots; exciton fatband plots; selected tests of the hybrid+$V_w$ approach with the SIESTA package

## REFERENCES

(1) Gibertini, M.; Koperski, M.; Morpurgo, A. F.; Novoselov, K. S. Magnetic 2D Materials and Heterostructures. *Nat. Nanotechnol.* **2019**, *14* (5), 408–419. https://doi.org/10.1038/s41565-019-0438-6.

(2) Mannix, A. J.; Kiraly, B.; Hersam, M. C.; Guisinger, N. P. Synthesis and Chemistry of Elemental 2D Materials. *Nat. Rev. Chem.* **2017**, *1* (2), 0014. https://doi.org/10.1038/s41570-016-0014.

(3) Huang, X.; Wang, T.; Miao, S.; Wang, C.; Li, Z.; Lian, Z.; Taniguchi, T.; Watanabe, K.; Okamoto, S.; Xiao, D.; Shi, S.-F.; Cui, Y.-T. Correlated Insulating States at Fractional Fillings of the WS2/WSe2 Moiré Lattice. *Nat. Phys.* **2021**, *17* (6), 715–719. https://doi.org/10.1038/s41567-021-01171-w.

(4) Cao, Y.; Fatemi, V.; Fang, S.; Watanabe, K.; Taniguchi, T.; Kaxiras, E.; Jarillo-Herrero, P. Unconventional Superconductivity in Magic-Angle Graphene Superlattices. *Nature* **2018**, *556* (7699), 43–50. https://doi.org/10.1038/nature26160.

(5) Xia, F.; Wang, H.; Xiao, D.; Dubey, M.; Ramasubramaniam, A. Two-Dimensional Material Nanophotonics. *Nat. Photonics* **2014**, *8* (12), 899–907.

(6) Zhou, X.; Hu, X.; Yu, J.; Liu, S.; Shu, Z.; Zhang, Q.; Li, H.; Ma, Y.; Xu, H.; Zhai, T. 2D Layered Material-Based van Der Waals Heterostructures for Optoelectronics. *Adv. Funct. Mater.* **2018**, *28* (14), 1706587. https://doi.org/10.1002/adfm.201706587.

(7) Lin, Z.; McCreary, A.; Briggs, N.; Subramanian, S.; Zhang, K.; Sun, Y.; Li, X.; Borys, N. J.; Yuan, H.; Fullerton-Shirey, S. K.; Chernikov, A.; Zhao, H.; McDonnell, S.; Lindenberg, A. M.; Xiao, K.; LeRoy, B. J.; Drndić, M.; Hwang, J. C. M.; Park, J.; Chhowalla, M.; Schaak, R. E.; Javey, A.; Hersam, M. C.; Robinson, J.; Terrones, M. 2D Materials Advances: From Large Scale Synthesis and Controlled Heterostructures to Improved Characterization Techniques, Defects and Applications. *2D Mater.* **2016**, *3* (4), 042001. https://doi.org/10.1088/2053-1583/3/4/042001.

(8) Dirnberger, F.; Bushati, R.; Datta, B.; Kumar, A.; MacDonald, A. H.; Baldini, E.; Menon, V. M. Spin-Correlated Exciton–Polaritons in a van Der Waals Magnet. *Nat. Nanotechnol.* **2022**, *17* (10), 1060–1064. https://doi.org/10.1038/s41565-022-01204-2.

(9) Lachance-Quirion, D.; Tabuchi, Y.; Gloppe, A.; Usami, K.; Nakamura, Y. Hybrid Quantum Systems Based on Magnonics. *Appl. Phys. Express* **2019**, *12* (7), 070101. https://doi.org/10.7567/1882-0786/ab248d.

(10) Zhang, X.; Zou, C.-L.; Jiang, L.; Tang, H. X. Strongly Coupled Magnons and Cavity Microwave Photons. *Phys. Rev. Lett.* **2014**, *113* (15), 156401. https://doi.org/10.1103/PhysRevLett.113.156401.

(11) Tabuchi, Y.; Ishino, S.; Noguchi, A.; Ishikawa, T.; Yamazaki, R.; Usami, K.; Nakamura, Y. Coherent Coupling between a Ferromagnetic Magnon and a Superconducting Qubit. *Science* **2015**, *349* (6246), 405–408. https://doi.org/10.1126/science.aaa3693.

(12) Bae, Y. J.; Wang, J.; Scheie, A.; Xu, J.; Chica, D. G.; Diederich, G. M.; Cenker, J.; Ziebel, M. E.; Bai, Y.; Ren, H.; Dean, C. R.; Delor, M.; Xu, X.; Roy, X.; Kent, A. D.; Zhu, X. Exciton-Coupled Coherent Magnons in a 2D Semiconductor. *Nature* **2022**, *609* (7926), 282–286. https://doi.org/10.1038/s41586-022-05024-1.

(13) Belvin, C. A.; Baldini, E.; Ozel, I. O.; Mao, D.; Po, H. C.; Allington, C. J.; Son, S.; Kim, B. H.; Kim, J.; Hwang, I.; Kim, J. H.; Park, J.-G.; Senthil, T.; Gedik, N. Exciton-Driven Antiferromagnetic Metal in a Correlated van Der Waals Insulator. *Nat. Commun.* **2021**, *12* (1), 4837. https://doi.org/10.1038/s41467-021-25164-8.

(14) Yuan, H. Y.; Cao, Y.; Kamra, A.; Duine, R. A.; Yan, P. Quantum Magnonics: When Magnon Spintronics Meets Quantum Information Science. *Quantum Magnonics Magnon Spintron. Meets Quantum Inf. Sci.* **2022**, *965*, 1–74. https://doi.org/10.1016/j.physrep.2022.03.002.

(15) Diederich, G. M.; Cenker, J.; Ren, Y.; Fonseca, J.; Chica, D. G.; Bae, Y. J.; Zhu, X.; Roy, X.; Cao, T.; Xiao, D.; Xu, X. Tunable Interaction between Excitons and Hybridized Magnons in a Layered Semiconductor. *Nat. Nanotechnol.* **2023**, *18* (1), 23–28. https://doi.org/10.1038/s41565-022-01259-1.

(16) Wilson, N. P.; Lee, K.; Cenker, J.; Xie, K.; Dismukes, A. H.; Telford, E. J.; Fonseca, J.; Sivakumar, S.; Dean, C.; Cao, T.; Roy, X.; Xu, X.; Zhu, X. Interlayer Electronic Coupling on Demand in a 2D Magnetic Semiconductor. *Nat. Mater.* **2021**, *20* (12), 1657–1662. https://doi.org/10.1038/s41563-021-01070-8.

(17) Ziebel, M. E.; Feuer, M. L.; Cox, J.; Zhu, X.; Dean, C. R.; Roy, X. CrSBr: An Air-Stable, Two-Dimensional Magnetic Semiconductor. *Nano Lett.* **2024**. https://doi.org/10.1021/acs.nanolett.4c00624.

(18) Marques-Moros, F.; Boix-Constant, C.; Mañas-Valero, S.; Canet-Ferrer, J.; Coronado, E. Interplay between Optical Emission and Magnetism in the van Der Waals Magnetic Semiconductor CrSBr in the Two-Dimensional Limit. *ACS Nano* **2023**, *17* (14), 13224–13231. https://doi.org/10.1021/acsnano.3c00375.

(19) Tabataba-Vakili, F.; Nguyen, H. P. G.; Rupp, A.; Mosina, K.; Papavasileiou, A.; Watanabe, K.; Taniguchi, T.; Maletinsky, P.; Glazov, M. M.; Sofer, Z.; Baimuratov, A. S.; Högele, A. Doping-Control of Excitons and Magnetism in Few-Layer CrSBr. *Nat. Commun.* **2024**, *15* (1), 4735. https://doi.org/10.1038/s41467-024-49048-9.

(20) Wang, T.; Zhang, D.; Yang, S.; Lin, Z.; Chen, Q.; Yang, J.; Gong, Q.; Chen, Z.; Ye, Y.; Liu, W. Magnetically-Dressed CrSBr Exciton-Polaritons in Ultrastrong Coupling Regime. *Nat. Commun.* **2023**, *14* (1), 5966. https://doi.org/10.1038/s41467-023-41688-7.

(21) Li, Q.; Xie, X.; Alfrey, A.; Beach, C. W.; McLellan, N.; Lu, Y.; Hu, J.; Liu, W.; Dhale, N.; Lv, B.; Zhao, L.; Sun, K.; Deng, H. Two-Dimensional Magnetic Exciton Polariton with Strongly Coupled Atomic and Photonic Anisotropies. *Phys. Rev. Lett.* **2024**, *133* (26), 266901. https://doi.org/10.1103/PhysRevLett.133.266901.

(22) Dirnberger, F.; Quan, J.; Bushati, R.; Diederich, G. M.; Florian, M.; Klein, J.; Mosina, K.; Sofer, Z.; Xu, X.; Kamra, A.; García-Vidal, F. J.; Alù, A.; Menon, V. M. Magneto-Optics in a van Der Waals Magnet Tuned by Self-Hybridized Polaritons. *Nature* **2023**, *620* (7974), 533–537. https://doi.org/10.1038/s41586-023-06275-2.

(23) Long, F.; Ghorbani-Asl, M.; Mosina, K.; Li, Y.; Lin, K.; Ganss, F.; Hübner, R.; Sofer, Z.; Dirnberger, F.; Kamra, A.; Krasheninnikov, A. V.; Prucnal, S.; Helm, M.; Zhou, S. Ferromagnetic Interlayer Coupling in CrSBr Crystals Irradiated by Ions. *Nano Lett.* **2023**, *23* (18), 8468–8473. https://doi.org/10.1021/acs.nanolett.3c01920.

(24) Göser, O.; Paul, W.; Kahle, H. G. Magnetic Properties of CrSBr. *J. Magn. Magn. Mater.* **1990**, *92* (1), 129–136. https://doi.org/10.1016/0304-8853(90)90689-N.

(25) López-Paz, S. A.; Guguchia, Z.; Pomjakushin, V. Y.; Witteveen, C.; Cervellino, A.; Luetkens, H.; Casati, N.; Morpurgo, A. F.; von Rohr, F. O. Dynamic Magnetic Crossover at the Origin of the Hidden-Order in van Der Waals Antiferromagnet CrSBr. *Nat. Commun.* **2022**, *13* (1), 4745. https://doi.org/10.1038/s41467-022-32290-4.

(26) Bianchi, M.; Acharya, S.; Dirnberger, F.; Klein, J.; Pashov, D.; Mosina, K.; Sofer, Z.; Rudenko, A. N.; Katsnelson, M. I.; van Schilfgaarde, M.; Rösner, M.; Hofmann, P. Paramagnetic Electronic Structure of CrSBr: Comparison between Ab Initio $GW$ Theory and Angle-Resolved Photoemission Spectroscopy. *Phys. Rev. B* **2023**, *107* (23), 235107. https://doi.org/10.1103/PhysRevB.107.235107.

(27) Watson, M. D.; Acharya, S.; Nunn, J. E.; Nagireddy, L.; Pashov, D.; Rösner, M.; van Schilfgaarde, M.; Wilson, N. R.; Cacho, C. Giant Exchange Splitting in the Electronic Structure of A-Type 2D Antiferromagnet CrSBr. *Npj 2D Mater. Appl.* **2024**, *8* (1), 1–8. https://doi.org/10.1038/s41699-024-00492-7.

(28) Klein, J.; Pingault, B.; Florian, M.; Heißenbüttel, M.-C.; Steinhoff, A.; Song, Z.; Torres, K.; Dirnberger, F.; Curtis, J. B.; Weile, M.; Penn, A.; Deilmann, T.; Dana, R.; Bushati, R.; Quan, J.; Luxa, J.; Sofer, Z.; Alù, A.; Menon, V. M.; Wurstbauer, U.; Rohlfing, M.; Narang, P.; Lončar, M.; Ross, F. M. The Bulk van Der Waals Layered Magnet CrSBr Is a Quasi-1D Material. *ACS Nano* **2023**, *17* (6), 5316–5328. https://doi.org/10.1021/acsnano.2c07316.

(29) Nessi, L.; Occhialini, C. A.; Demir, A. K.; Powalla, L.; Comin, R. Magnetic Field Tunable Polaritons in the Ultrastrong Coupling Regime in CrSBr. *ACS Nano* **2024**, *18* (50), 34235–34243. https://doi.org/10.1021/acsnano.4c11799.

(30) Klein, J.; Ross, F. M. Materials beyond Monolayers: The Magnetic Quasi-1D Semiconductor CrSBr. *J. Mater. Res.* **2024**, *39* (22), 3045–3056. https://doi.org/10.1557/s43578-024-01459-6.

(31) Smolenski, S.; Wen, M.; Li, Q.; Downey, E.; Alfrey, A.; Liu, W.; Kondusamy, A. L. N.; Bostwick, A.; Jozwiak, C.; Rotenberg, E.; Zhao, L.; Deng, H.; Lv, B.; Zgid, D.; Gull, E.; Jo, N. H. Large Exciton Binding Energy in a Bulk van Der Waals Magnet from Quasi-1D Electronic Localization. *Nat. Commun.* **2025**, *16* (1), 1134. https://doi.org/10.1038/s41467-025-56457-x.

(32) Shao, Y.; Dirnberger, F.; Qiu, S.; Acharya, S.; Terres, S.; Telford, E. J.; Pashov, D.; Kim, B. S. Y.; Ruta, F. L.; Chica, D. G.; Dismukes, A. H.; Ziebel, M. E.; Wang, Y.; Choe, J.; Bae, Y. J.; Millis, A. J.; Katsnelson, M. I.; Mosina, K.; Sofer, Z.; Huber, R.; Zhu, X.; Roy,

X.; van Schilfgaarde, M.; Chernikov, A.; Basov, D. N. Magnetically Confined Surface and Bulk Excitons in a Layered Antiferromagnet. *Nat. Mater.* **2025**, *24* (3), 391–398. https://doi.org/10.1038/s41563-025-02129-6.

(33) Ramasubramaniam, A. Large Excitonic Effects in Monolayers of Molybdenum and Tungsten Dichalcogenides. *Phys. Rev. B* **2012**, *86* (11), 115409. https://doi.org/10.1103/PhysRevB.86.115409.

(34) Camarasa-Gómez, M.; Ramasubramaniam, A.; Neaton, J. B.; Kronik, L. Transferable Screened Range-Separated Hybrid Functionals for Electronic and Optical Properties of van Der Waals Materials. *Phys. Rev. Mater.* **2023**, *7* (10), 104001. https://doi.org/10.1103/PhysRevMaterials.7.104001.

(35) Camarasa-Gómez, M.; Gant, S. E.; Ohad, G.; Neaton, J. B.; Ramasubramaniam, A.; Kronik, L. Excitations in Layered Materials from a Non-Empirical Wannier-Localized Optimally-Tuned Screened Range-Separated Hybrid Functional. *Npj Comput. Mater.* **2024**, *10* (1), 1–9. https://doi.org/10.1038/s41524-024-01478-1.

(36) Śmiertka, M.; Rygała, M.; Posmyk, K.; Peksa, P.; Dyksik, M.; Pashov, D.; Mosina, K.; Sofer, Z.; van Schilfgaarde, M.; Dirnberger, F.; Baranowski, M.; Acharya, S.; Plochocka, P. Distinct Magneto-Optical Response of Frenkel and Wannier Excitons in CrSBr. *Nat. Commun.* **2026**, *17* (1), 1777. https://doi.org/10.1038/s41467-026-68482-5.

(37) Klein, J.; Pham, T.; Thomsen, J. D.; Curtis, J. B.; Denneulin, T.; Lorke, M.; Florian, M.; Steinhoff, A.; Wiscons, R. A.; Luxa, J.; Sofer, Z.; Jahnke, F.; Narang, P.; Ross, F. M. Control of Structure and Spin Texture in the van Der Waals Layered Magnet CrSBr. *Nat. Commun.* **2022**, *13* (1), 5420. https://doi.org/10.1038/s41467-022-32737-8.

(38) Wang, Y.; Luo, N.; Zeng, J.; Tang, L.-M.; Chen, K.-Q. Magnetic Anisotropy and Electric Field Induced Magnetic Phase Transition in the van Der Waals Antiferromagnet CrSBr. *Phys. Rev. B* **2023**, *108* (5), 054401. https://doi.org/10.1103/PhysRevB.108.054401.

(39) Liu, J.; Zhang, X.; Lu, G. Moiré Magnetism and Moiré Excitons in Twisted CrSBr Bilayers. *Proc. Natl. Acad. Sci.* **2025**, *122* (1), e2413326121. https://doi.org/10.1073/pnas.2413326121.

(40) Feuer, M. L.; Thinel, M.; Huang, X.; Cui, Z.-H.; Shao, Y.; Kundu, A. K.; Chica, D. G.; Han, M.-G.; Pokratath, R.; Telford, E. J.; Cox, J.; York, E.; Okuno, S.; Huang, C.-Y.; Bukula, O.; Nashabeh, L. M.; Qiu, S.; Nuckolls, C. P.; Dean, C. R.; Billinge, S. J. L.; Zhu,

X.; Zhu, Y.; Basov, D. N.; Millis, A. J.; Reichman, D. R.; Pasupathy, A. N.; Roy, X.; Ziebel, M. E. Charge Density Wave and Ferromagnetism in Intercalated CrSBr. *Adv. Mater.* **2025**, *37* (24), 2418066. https://doi.org/10.1002/adma.202418066.

(41) Esteras, D. L.; Rybakov, A.; Ruiz, A. M.; Baldoví, J. J. Magnon Straintronics in the 2D van Der Waals Ferromagnet CrSBr from First-Principles. *Nano Lett.* **2022**, *22* (21), 8771–8778. https://doi.org/10.1021/acs.nanolett.2c02863.

(42) Heißenbüttel, M.-C.; Piel, P.-M.; Klein, J.; Deilmann, T.; Wurstbauer, U.; Rohlfing, M. Quadratic Optical Response to a Magnetic Field in the Layered Magnet CrSBr. *Phys. Rev. B* **2025**, *111* (7), 075107. https://doi.org/10.1103/PhysRevB.111.075107.

(43) Kalitukha, I. V.; Akimov, I. A.; Nestoklon, M. O.; Geirsson, T.; Molina-Sánchez, A.; Yalcin, E.; Ruppert, C.; Mayoh, D. A.; Balakrishnan, G.; Karuppasamy, M.; Sofer, Z.; Wang, Y.; Gillard, D. J.; Hu, X.; Tartakovskii, A. I.; Bayer, M. Magnetic Switching of Exciton Lifetime in CrSBr. arXiv January 8, 2026. https://doi.org/10.48550/arXiv.2601.05413.

(44) Wu, F.; Zhang, X.; Chen, Y.; Pei, D.; Zhan, M.; Tao, Z.; Chen, C.; Lu, S.; Chen, J.; Tang, S.; Wang, X.; Guo, Y.; Yang, L.; Zhang, Y.; Chen, Y.; Mi, Q.; Li, G.; Liu, Z. Mott Insulating Phase and Coherent-Incoherent Crossover across Magnetic Phase Transition in 2D Antiferromagnetic CrSBr. *Sci. China Phys. Mech. Astron.* **2025**, *68* (6), 267411. https://doi.org/10.1007/s11433-025-2625-7.

(45) Porée, V.; Zobelli, A.; Pawbake, A.; Regner, J.; Sofer, Z.; Faugeras, C.; Nicolaou, A. Resonant X-Ray Spectroscopies on CrSBr: Probing the Electronic Structure through Chromium d–d Excitations. *Phys. Rev. B* **2025**, *112* (12), 125103. https://doi.org/10.1103/pdvz-6cpg.

(46) Sears, J.; Zager, B.; He, W.; Occhialini, C. A.; Shen, Y.; Lajer, M.; Villanova, J. W.; Berlijn, T.; Yakhou-Harris, F.; Brookes, N. B.; Chica, D. G.; Roy, X.; Baldini, E.; Pelliciari, J.; Bisogni, V.; Johnston, S.; Mitrano, M.; Dean, M. P. M. Observation of Anisotropic Dispersive Dark-Exciton Dynamics in CrSBr. *Phys. Rev. Lett.* **2025**, *135* (14), 146503. https://doi.org/10.1103/fz3h-6jdx.

(47) Perdew, J. P.; Burke, K.; Ernzerhof, M. Generalized Gradient Approximation Made Simple. *Phys. Rev. Lett.* **1996**, *77* (18), 3865–3868. https://doi.org/10.1103/PhysRevLett.77.3865.

(48) Bartók, A. P.; Yates, J. R. Regularized SCAN Functional. *J. Chem. Phys.* **2019**, *150* (16), 161101. https://doi.org/10.1063/1.5094646.

(49) Krukau, A. V.; Vydrov, O. A.; Izmaylov, A. F.; Scuseria, G. E. Influence of the Exchange Screening Parameter on the Performance of Screened Hybrid Functionals. *J. Chem. Phys.* **2006**, *125* (22), 224106. https://doi.org/10.1063/1.2404663.

(50) Anisimov, V. I.; Aryasetiawan, F.; Lichtenstein, A. I. First-Principles Calculations of the Electronic Structure and Spectra of Strongly Correlated Systems: The **LDA** + *U* Method. *J. Phys. Condens. Matter* **1997**, *9* (4), 767–808. https://doi.org/10.1088/0953-8984/9/4/002.

(51) Himmetoglu, B.; Floris, A.; de Gironcoli, S.; Cococcioni, M. Hubbard-Corrected DFT Energy Functionals: The LDA+U Description of Correlated Systems. *Int. J. Quantum Chem.* **2014**, *114* (1), 14–49. https://doi.org/10.1002/qua.24521.

(52) Macke, E.; Timrov, I.; Marzari, N.; Ciacchi, L. C. Orbital-Resolved DFT+U for Molecules and Solids. *J. Chem. Theory Comput.* **2024**, *20* (11), 4824–4843. https://doi.org/10.1021/acs.jctc.3c01403.

(53) Yang, Z.; Ullrich, C. A. Direct Calculation of Exciton Binding Energies with Time-Dependent Density-Functional Theory. *Phys. Rev. B* **2013**, *87* (19), 195204. https://doi.org/10.1103/PhysRevB.87.195204.

(54) Yang, Z.; Sottile, F.; Ullrich, C. A. Simple Screened Exact-Exchange Approach for Excitonic Properties in Solids. *Phys. Rev. B* **2015**, *92* (3), 035202. https://doi.org/10.1103/PhysRevB.92.035202.

(55) Wing, D.; Ohad, G.; Haber, J. B.; Filip, M. R.; Gant, S. E.; Neaton, J. B.; Kronik, L. Band Gaps of Crystalline Solids from Wannier-Localization–Based Optimal Tuning of a Screened Range-Separated Hybrid Functional. *Proc. Natl. Acad. Sci.* **2021**, *118* (34), e2104556118. https://doi.org/10.1073/pnas.2104556118.

(56) Wing, D.; Haber, J. B.; Noff, R.; Barker, B.; Egger, D. A.; Ramasubramaniam, A.; Louie, S. G.; Neaton, J. B.; Kronik, L. Comparing Time-Dependent Density Functional Theory with Many-Body Perturbation Theory for Semiconductors: Screened Range-Separated Hybrids and the $GW$ plus Bethe-Salpeter Approach. **2019**, 1–15. https://doi.org/10.1103/PhysRevMaterials.3.064603.

(57) Ramasubramaniam, A.; Wing, D.; Kronik, L. Transferable Screened Range-Separated Hybrids for Layered Materials: The Cases of MoS2 and h-BN. *Phys. Rev. Mater.* **2019**, *3* (8), 084007.

(58) Ivády, V.; Abrikosov, I. A.; Janzén, E.; Gali, A. Role of Screening in the Density Functional Applied to Transition-Metal Defects in Semiconductors. *Phys. Rev. B* **2013**, *87* (20), 205201. https://doi.org/10.1103/PhysRevB.87.205201.

(59) Ivády, V.; Armiento, R.; Szász, K.; Janzén, E.; Gali, A.; Abrikosov, I. A. Theoretical Unification of Hybrid-DFT and $\text{DFT} + U$ Methods for the Treatment of Localized Orbitals. *Phys. Rev. B* **2014**, *90* (3), 035146. https://doi.org/10.1103/PhysRevB.90.035146.

(60) Ivády, V.; Gali, A.; Abrikosov, I. A. Hybrid-DFT + $V_w$ Method for Band Structure Calculation of Semiconducting Transition Metal Compounds: The Case of Cerium Dioxide. *J. Phys. Condens. Matter* **2017**, *29* (45), 454002. https://doi.org/10.1088/1361-648X/aa8b93.

(61) Dudarev, S. L.; Botton, G. A.; Savrasov, S. Y.; Humphreys, C. J.; Sutton, A. P. Electron-Energy-Loss Spectra and the Structural Stability of Nickel Oxide: An LSDA+U Study. *Phys. Rev. B* **1998**, *57* (3), 1505. https://doi.org/10.1103/PhysRevB.57.1505.

(62) Wu, F.; Gutiérrez-Lezama, I.; López-Paz, S. A.; Gibertini, M.; Watanabe, K.; Taniguchi, T.; von Rohr, F. O.; Ubrig, N.; Morpurgo, A. F. Quasi-1D Electronic Transport in a 2D Magnetic Semiconductor. *Adv. Mater.* **2022**, *34* (16), 2109759. https://doi.org/10.1002/adma.202109759.

(63) Lloyd, L. T.; Pincelli, T.; Wahada, M. A.; Vita, A. D.; Menzel, F.; Mosina, K.; Castro, T. H. L. G.; Neef, A.; Stier, A. V.; Wilson, N. P.; Sofer, Z.; Finley, J. J.; Wolf, M.; Rettig, L.; Ernstorfer, R. Ultrafast Formation and Annihilation of Strongly Bound, Anisotropic Excitons, 2026. https://arxiv.org/abs/2603.26294.

(64) Nguyen, N. L.; Colonna, N.; Ferretti, A.; Marzari, N. Koopmans-Compliant Spectral Functionals for Extended Systems. *Phys. Rev. X* **2018**, *8* (2), 021051. https://doi.org/10.1103/PhysRevX.8.021051.

(65) Ma, J.; Wang, L.-W. Using Wannier Functions to Improve Solid Band Gap Predictions in Density Functional Theory. *Sci. Rep.* **2016**, *6* (1), 24924. https://doi.org/10.1038/srep24924.

(66) Ohad, G.; Camarasa-Gómez, M.; Neaton, J. B.; Ramasubramaniam, A.; Gould, T.; Kronik, L. Foundations of the Ionization Potential Condition for Localized Electron Removal in Density Functional Theory. *Phys. Rev. Lett.* **2026**, *136* (2), 026401. https://doi.org/10.1103/ngyq-q7z3.

(67) Kresse, G.; Furthmüller, J. Efficient Iterative Schemes for Ab Initio Total-Energy Calculations Using a Plane-Wave Basis Set. *Phys. Rev. B* **1996**, *54* (16), 11169–11186. https://doi.org/10.1103/PhysRevB.54.11169.

(68) Kresse, G.; Furthmüller, J. Efficiency of Ab-Initio Total Energy Calculations for Metals and Semiconductors Using a Plane-Wave Basis Set. *Comput. Mater. Sci.* **1996**, *6* (1), 15–50. https://doi.org/10.1016/0927-0256(96)00008-0.

(69) Kresse, G.; Joubert, D. From Ultrasoft Pseudopotentials to the Projector Augmented-Wave Method. *Phys. Rev. B* **1999**, *59* (3), 1758–1775. https://doi.org/10.1103/PhysRevB.59.1758.

(70) Casida, M. E. Time-Dependent Density Functional Response Theory for Molecules. In *Recent Advances in Density Functional Methods*; Recent Advances in Computational Chemistry; WORLD SCIENTIFIC, 1995; Vol. Volume 1, pp 155–192. https://doi.org/10.1142/9789812830586_0005.

(71) Onida, G.; Reining, L.; Rubio, A. Electronic Excitations: Density-Functional versus Many-Body Green's-Function Approaches. *Rev. Mod. Phys.* **2002**, *74* (2), 601–659. https://doi.org/10.1103/RevModPhys.74.601.

# SUPPLEMENTARY INFORMATION

## Efficient prediction of highly anisotropic excitonic properties in the layered antiferromagnet CrSBr via time-dependent density functional theory

**Ashwin Ramasubramaniam,**[1,2,*] **Daniel Hernangómez-Pérez,**[3] **Javier Junquera,**[4] **María Camarasa-Gómez**[5]

[1] Department of Mechanical and Industrial Engineering, University of Massachusetts, Amherst, MA 01003, U.S.A.

[2] Graduate Program in Materials Science and Engineering, University of Massachusetts, Amherst, MA 01003, U.S.A.

[3] CIC nanoGUNE BRTA, Tolosa Hiribidea 76, 20018 San Sebastián, Spain

[4] Departamento de Ciencias de la Tierra y Física de la Materia Condensada, Facultad de Ciencias, Universidad de Cantabria, Avenida de los Castros s/n, E-39005 Santander, Spain

[5] Centro de Física de Materiales (CFM-MPC), CSIC-UPV/EHU, Paseo Manuel de Lardizabal 5 20018 Donostia-San Sebastián, Spain

**Corresponding Author**

* ashwin@engin.umass.edu

## S1. SUPPORTING FIGURES FOR GLOBAL AND RANGE-SEPARATED HYBRIDS

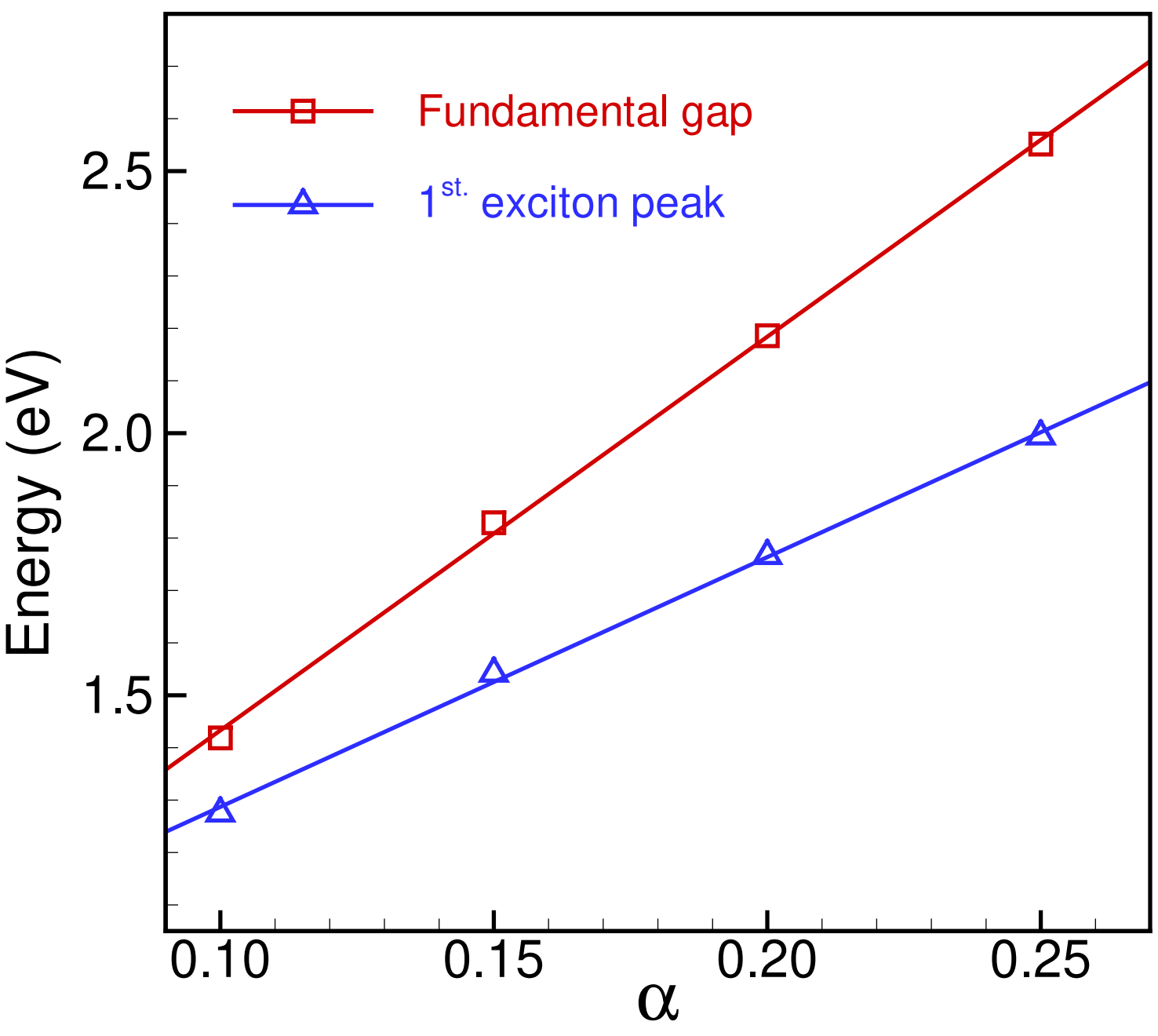

**Figure S1:** Fundamental gap and energy of first exciton peak as a function of global fraction of exact exchange, α.

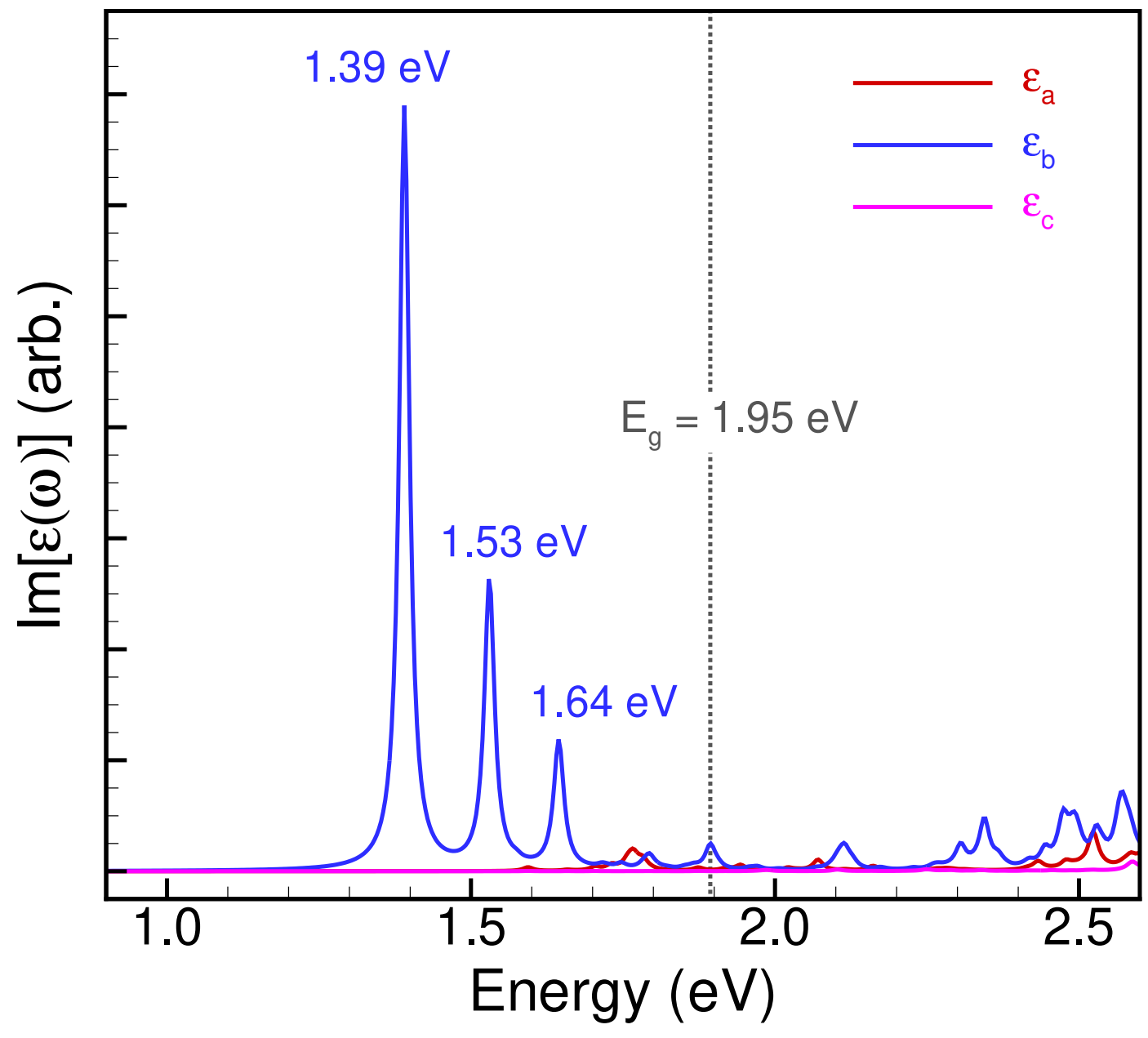

**Figure S2:** Optical spectrum (imaginary part of frequency-dependent dielectric function) for a range-separated hybrid with short-range fraction of exact exchange α=0.1, long-range fraction of exact-exchange β=0.3, and range-separation parameter γ=0.2 $Å^{-1}$.

## S2. UNIT CELL AND ATOMIC COORDINATES (VASP POSCAR)

Atomic positions and lattice parameters from Ref. 1.

```
Cr S Br AFM bulk
1.0
        3.5120599270         0.0000000000         0.0000000000
        0.0000000000         4.7447900772         0.0000000000
        0.0000000000         0.0000000000        15.8240995407
   Cr    S   Br
    4    4    4
Direct
0.750000000  0.250000000  0.564059973
0.750000000  0.250000000  0.064059973
0.250000000  0.750000000  0.435940027
0.250000000  0.750000000  0.935940027
0.750000000  0.750000000  0.538500011
0.750000000  0.750000000  0.038500011
0.250000000  0.250000000  0.461499989
0.250000000  0.250000000  0.961499989
0.750000000  0.750000000  0.323599994
0.750000000  0.750000000  0.823599994
0.250000000  0.250000000  0.676400006
0.250000000  0.250000000  0.176400006
```

## S3. PARAMETRIC DEPENDENCE OF FUNDAMENTAL GAPS AND EXCITON ENERGIES ON α AND $V_W$

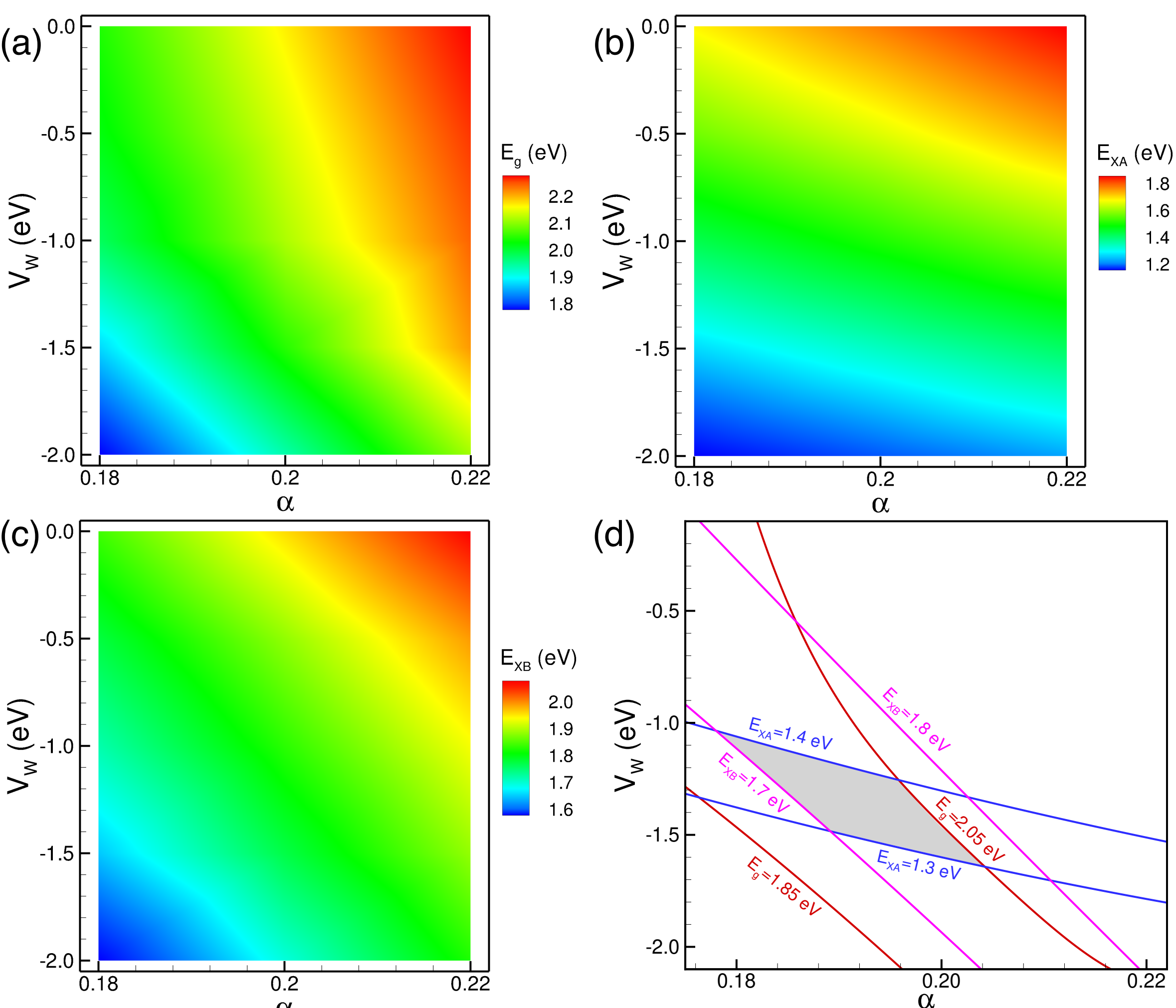

**Figure S3:** Contour plots depicting the dependence of (a) the fundamental gap, $E_g$, (b) the $X_A$ exciton energy, and (c) the $X_B$ exciton energy on the fraction of global exact exchange α and the onsite correction, $V_w$, applied to Cr *d*-orbitals. (d) Contours of $E_g$ (red lines), $X_A$ (blue lines), and $X_B$ (magenta lines) superposed to find the space of feasible solutions (grey region) for α and $V_w$. The limiting values of $E_g$ are 1.85 eV and 2.05 eV (as explained in the main text); for the $X_A$ and $X_B$ exciton energies, we use narrow 10-meV windows of uncertainty ranging from 1.3-1.4 eV and 1.7-1.8 eV, respectively.

The two independent parameters α and $V_w$ can be determined efficiently by performing a parameter sweep in α–$V_w$ space and recording the fundamental gap ($E_g$) as well as the $X_A$ and $X_B$ exciton energies. Figures S3(a-c) display these data for $E_g$, and $X_A$ and $X_B$ exciton energies, which can then be combined with known bounds from the literature to identify a space of feasible solutions [Fig. S3(d)]. Interestingly, if we reply purely on the energies of the $X_A$ and $X_B$ excitons – to which we apply a small 10-meV window of uncertainty – the fundamental gap is consistently predicted to be above the ARPES lower bound of 1.85 eV. At the upper end, the fundamental gap is bounded by 2.10 eV, which is only slightly above the theoretical $QSG\hat{W}$ value of ~2.05 eV. If we enforce precise values of 1.36 eV and 1.76 eV for the $X_A$ and $X_B$ excitons, respectively, the fundamental gap is predicted to be 2.02 eV, within expected bounds.

The parameter sweeps displayed here were carried out without spin-orbit coupling; time-dependent calculations of the optical spectrum were carried out using 16 occupied and 16 empty states, which is sufficient to converge the energy of the $X_B$ exciton to within 5 meV and that of the $X_A$ exciton to ~40 meV. Having identified a reasonable window of parameter space, more targeted and computationally expensive calculations with spin-orbit coupling can be performed to identify a suitable $\alpha$ and $V_w$, selected results for which are displayed in Figure 2 of the main article.

It is noteworthy that the tuning process can rely almost entirely on high-quality optical measurements at low temperatures that are much less involved than low-temperature ARPES measurements.

## S4. EXCITON FATBAND PLOTS

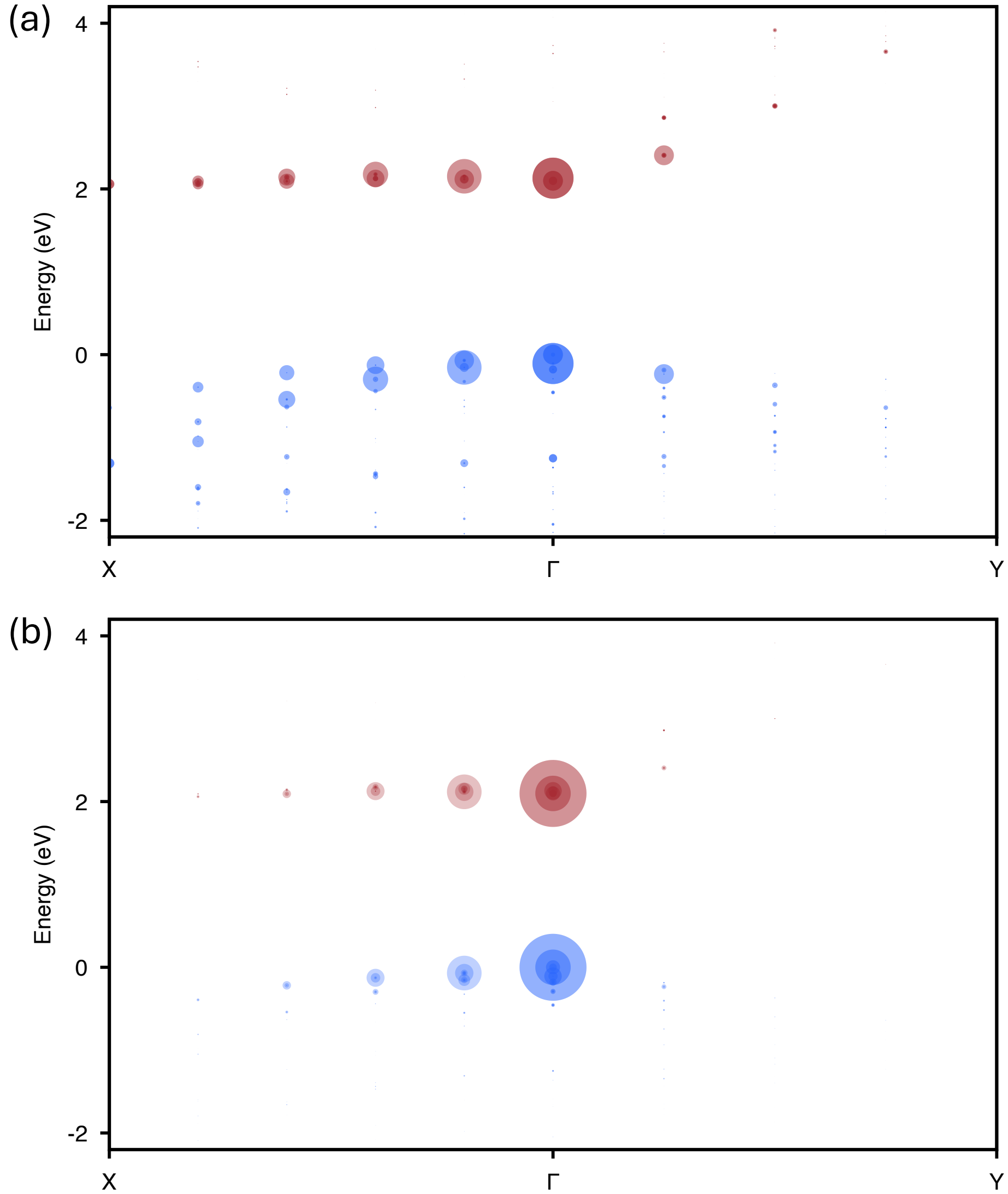

**Figure S4:** Fatband plots for the (a) $X_A$ and (b) $X_B$ excitons of the bulk AFM phase. Blue and red circles indicate valence and conduction states with their radii representing the coupling strength. The valence band edge energy is set to zero. Fatband plots are converted to heatmaps (Figure 3 of manuscript) by a standard rasterization process, using a 500×700 *k*-point–energy grid and a Gaussian smoothing of 0.02 in both dimensions to produce a 2D intensity map that is finally normalized (between 0 and 1) and overlaid on the bandstructure.

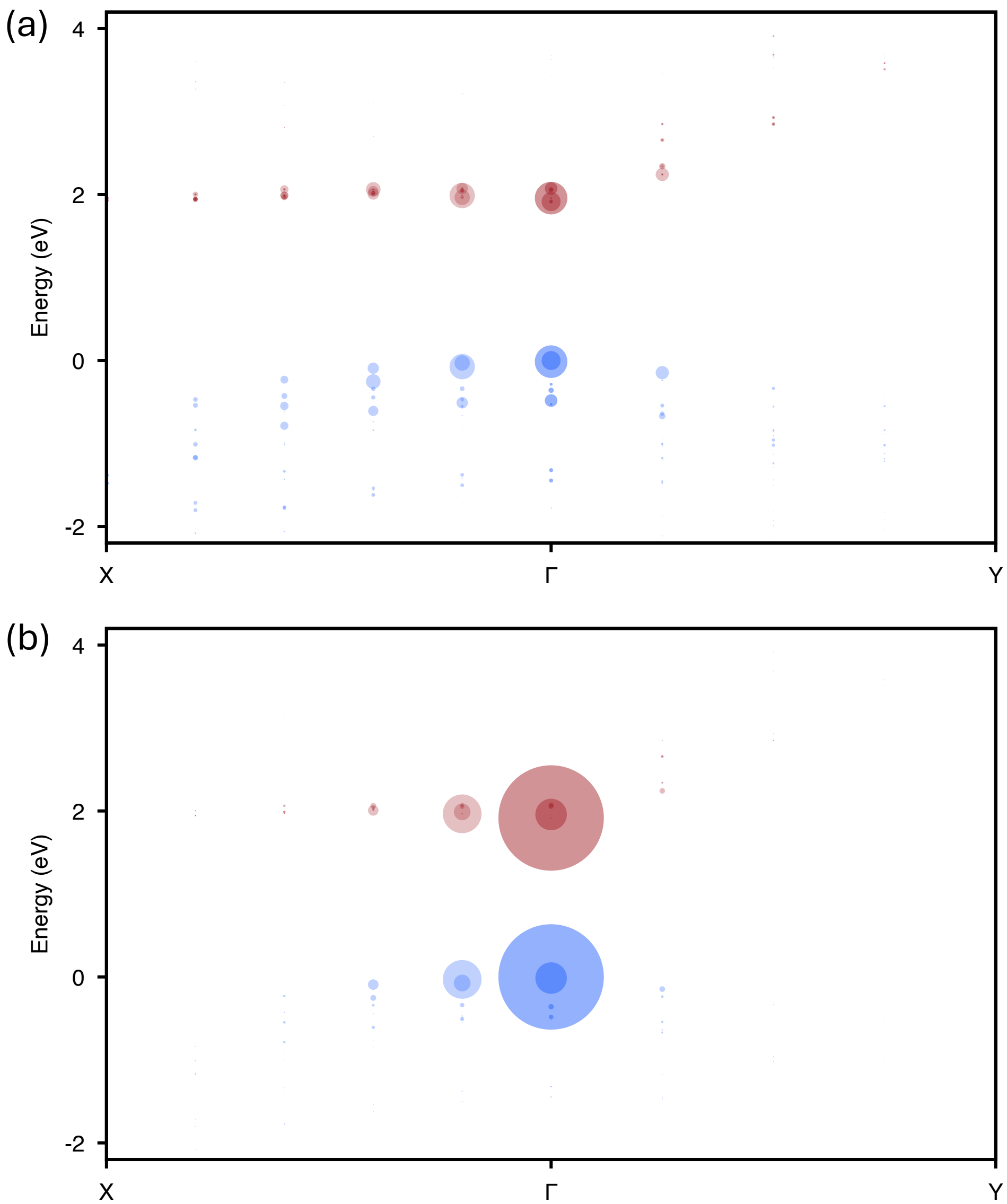

**Figure S5:** Same as Figure S4 except for the c-FM bulk phase.

## S5. HYBRID+$V_W$ TESTS WITH SIESTA

To confirm that conclusions drawn from the hybrid+$V_w$ calculations in VASP are not a result of artifacts, we performed a complementary set of selected tests with another first-principles code SIESTA. The SIESTA calculations were performed using the PBE0 hybrid functional, employing the standard 25% fraction of Hartree–Fock exchange.[2] The implementation of this functional within the numerical atomic-orbital framework of SIESTA has been carried out only recently.[3,4]

Core electrons were replaced by *ab initio* norm-conserving fully separable pseudopotentials.[5] In this work the optimized norm-conserving Vanderbilt pseudopotentials proposed by Hamann [6] were used, in the PSML[7] format available in the Pseudo-Dojo periodic table.[8,9] For Cr, the semicore 3*s* and 3*p* electrons were explicitly included in the valence. For Br, the semicore 3*d* shell was also considered. For S, the valence configuration was made of the 3*s* and 3*p* orbitals. In the generation of the pseudopotentials, the exchange-correlation functional was approximated using the Perdew-Burke-Ernzerhof (PBE) functional.[10]

The one-electron Kohn-Sham eigenstates were expanded in a basis of strictly localized [11] numerical atomic orbitals.[12] The native basis functions were obtained by finding the eigenfunctions of the isolated atoms confined within the soft-confinement spherical potential proposed in Ref. 13. A single-$\zeta$ basis set was applied to the 3*s* and 3*p* semicore states of Cr, and for the 3*d* semicore states of Br. For the valence states of Cr, S, and Br, we used a basis of double-$\zeta$ polarized quality, with all the parameters controlling the range and the shape taken from default. In the hybrid-functional implementation of SIESTA, the native numerical atomic orbitals are replaced by a fitted expansion in Gaussian functions. The maximum number of Gaussians in this expansion was set to four, and the contracted sum was truncated at the radius where $r^l$ times the Gaussian combination falls below $5\times10^{-3}$. The threshold used to screen four-center integrals via the Schwarz inequality was fixed to $1\times10^{-6}$. In addition to the hybrid functional, we included an onsite correction on the Cr 3*d* manifold following the Dudarev formalism.[14] The corresponding atomic projectors were generated using the same procedure employed for the construction of the native basis set, enforcing a tight cutoff radius of 2.5 Bohr.

The electronic density, Hartree, and exchange-correlation potentials, as well as the corresponding matrix elements between the basis orbitals, were calculated in a uniform real space grid.[16] An equivalent plane-wave cutoff of 1200 Ry was used to represent the charge density. The integrals in reciprocal space were well converged, using in all the cases a sampling in reciprocal space of the same quality as the $(10\times8\times2)$ Monkhorst-Pack mesh.[16] Atomic coordinates and lattice vectors used were the same as those for VASP (Sec. S2).

Figure S6 displays the projected density of states for three cases with $V_w = 0$ eV (no onsite correction) and $V_w = \pm1$ eV. Similar to the results in Figure 2 of the main manuscript, we find that for zero or positive $V_w$ the occupied Cr levels are too deep in the valence band, with the band edge being dominated by anion states. With negative values of $V_w$, the occupied Cr levels shift towards the valence band edge, and the relative contribution of anion states becomes smaller.

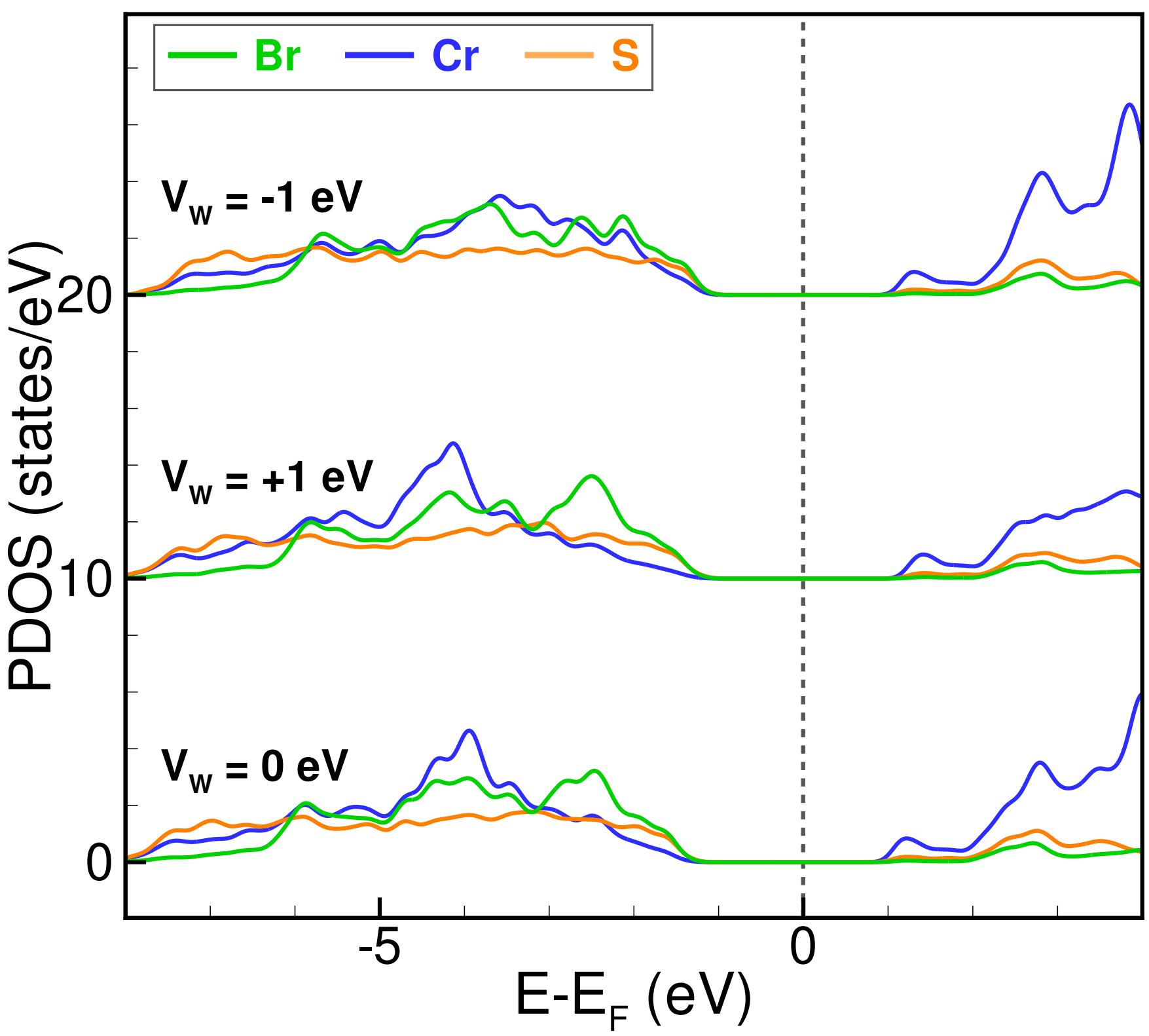

**Figure S6:** Species-wise PDOS of bulk CrSBr using the hybrid+Vw approach in SIESTA with the PBE0 (a=0.25) global hybrid. The Fermi level is set mid-gap and only the majority spin PDOS are indicated, the minority spin channel being degenerate.

**REFERENCES**

[1] S. A. López-Paz, Z. Guguchia, V. Y. Pomjakushin, C. Witteveen, A. Cervellino, H. Luetkens, N. Casati, A. F. Morpurgo, and F. O. von Rohr, Dynamic Magnetic Crossover at the Origin of the Hidden-Order in van Der Waals Antiferromagnet CrSBr. Nat. Commun. 13 (1), 4745 (2022).

[2] J. P. Perdew, M. Ernzerhof, and K. Burke, Rationale for mixing exact exchange with density functional approximations, J. Chem. Phys. 105, 9982 (1996).

[3] A. García, N. Papior, A. Akhtar, E. Artacho, V. Blum, E. Bosoni, P. Brandimarte, M. Brandbyge, J. I. Cerdá, F. Corsetti, R. Cuadrado, V. Dikan, J. Ferrer, J. Gale, P. García-Fernández, V. M. García-Suárez, S. García, G. Huhs, S. Illera, R. Korytár, P. Koval, I. Lebedeva, L. Lin, P. López-Tarifa, S. G. Mayo, S. Mohr, P. Ordejón, A. Postnikov, Y. Pouillon, M. Pruneda, R. Robles, D. Sánchez-Portal, J. M. Soler, R. Ullah, V. W.-Z. Yu, and J. Junquera, SIESTA: Recent developments and applications, J. Chem. Phys. 152, 204108 (2020).

[4] Y. Pouillon, B. C. Oyomo, J. Sifuna, M. Camarasa-Gómez, X. Qin, C. Beltrán, F. Gómez-Ortiz, H. Shang, and J. Junquera, Implementation of the hybrid exchange-correlation functionals in the SIESTA code, Comput. Phys. Commun. 323, 110086 (2026).

[5] L. Kleinman and D. M. Bylander, Efficacious form for model pseudopotentials, Phys. Rev. Lett. 48, 1425 (1982).

[6] D. Hamann, Optimized norm-conserving Vanderbilt pseudopotentials, Phys. Rev. B 88, 085117 (2013).

[7] A. García, M. J. Verstraete, Y. Pouillon, and J. Junquera, The PSML format and library for norm-conserving pseudopotential data curation and interoperability, Comput. Phys. Commun. 227, 51 (2018).

[8] M. Van Setten, M. Giantomassi, E. Bousquet, M. J. Verstraete, D. R. Hamann, X. Gonze, and G.-M. Rignanese, The PseudoDojo: Training and grading a 85 element optimized norm-conserving pseudopotential table, Comput. Phys. Commun. 226, 39 (2018).

[9] The scalar relativistic oncvpsp v0.4.1 pseudopotentials with stringent accuracy were used.

[10] J. P. Perdew, K. Burke, and M. Ernzerhof, Generalized gradient approximation made simple, Phys. Rev. Lett. 77, 3865 (1996).

[11] O. F. Sankey and D. J. Niklewski, Ab initio multicenter tight-binding model for molecular-dynamics simulations and other applications in covalent systems, Phys. Rev. B 40, 3979 (1989).

[12] E. Artacho, D. Sánchez-Portal, P. Ordejón, A. García, and J. M. Soler, Linear-scaling ab-initio calculations for large and complex systems, Phys. Stat. Sol. (b) 215, 809 (1999).

[13] J. Junquera, O. Paz, D. Sánchez-Portal, and E. Artacho, Numerical atomic orbitals for linear-scaling calculations, Phys. Rev. B 64, 235111 (2001).

[14] S. L. Dudarev, G. A. Botton, S. Y. Savrasov, C. J. Humphreys, and A. P. Sutton, Electron-energy-loss spec- tra and the structural stability of nickel oxide: An LSDA + U study, Phys. Rev. B 57, 1505 (1998).

[15] J. M. Soler, E. Artacho, J. D. Gale, A. García, J. Junquera, P. Ordejón, and D. Sánchez-Portal, The SIESTA method for ab initio order-N materials simulation, J. Phys.: Condens. Matter 14, 2745 (2002).

[16] H. J. Monkhorst and J. D. Pack, Special points for Brillouin-zone integrations, Phys. Rev. B 13, 5188 (1976).